\documentclass{article}
\usepackage[nohyperref,accepted]{icml2023}

\usepackage[compact]{titlesec}
\titlespacing*{\section}{0pt}{0pt}{0pt}
\titlespacing*{\subsection}{0pt}{0pt}{0pt}

\usepackage{float}  %
\usepackage{wrapfig}
\usepackage{caption}
\usepackage[utf8]{inputenc} %
\usepackage[T1]{fontenc}    %
\usepackage{hyperref}       %
\usepackage{url}            %
\usepackage{booktabs}       %
\usepackage{amsfonts}       %
\usepackage{nicefrac}       %
\usepackage{microtype}      %
\usepackage{amsmath}
\usepackage{subfigure}

\usepackage{forloop}
\usepackage{xspace}
\usepackage{graphicx}
\usepackage{mathtools}

\usepackage{thmtools}
\usepackage{thm-restate}
\usepackage{xcolor}
\usepackage{siunitx, etoolbox}
\usepackage{multirow}

\newcommand{\defvec}[1]{\expandafter\newcommand\csname v#1\endcsname{{\mathbf{#1}}}}
\newcounter{ct}
\forLoop{1}{26}{ct}{
	\edef\letter{\alph{ct}}
	\expandafter\defvec\letter
}
\forLoop{1}{26}{ct}{
	\edef\letter{\Alph{ct}}
	\expandafter\defvec\letter
}

\newcommand{\latents}{\vz}

\newcommand{\field}[1]{\ensuremath{\mathbb{#1}}}

\newcommand{\reals}{\field{R}}

\DeclareMathOperator*{\argmax}{\arg\!\max}
\DeclareMathOperator*{\argmin}{\arg\!\min}

\newcommand{\CVIHMGP}{cvHM\xspace}
\newcommand{\HidaMatern}{Hida-Mat\'ern\xspace}
\newcommand{\HM}{HM\xspace}

\newcommand{\Matern}{Mat\'ern\xspace}

\newcommand{\HidaMatrixDiff}{\vK^S}
\newcommand{\HidaStateTransition}{\vA}
\newcommand{\HidaStateNoise}{\boldsymbol{\epsilon}}

\newcommand{\VecZero}{\mathbf{0}}

\definecolor{retroblue1}{cmyk}{0.89, 0.46, 0.24, 0.04}
\definecolor{retroblue2}{cmyk}{0.58, 0.15, 0.40, 0}
\definecolor{retropale1}{cmyk}{0, 0.72, 0.91, 0}
\definecolor{mpcolor}{rgb}{1, 0.1, 0.59}

\newcommand{\numLatents}{L}

\newcommand{\priorHypers}{\boldsymbol{\theta}}

\newcommand{\naturalVariationalParams}{\boldsymbol{\lambda}}

\newcommand{\auxNaturalVariationalParams}{\tilde{\boldsymbol{\lambda}}}
\newcommand{\auxLikelihood}{\tilde{\vy}}
\newcommand{\auxParamOne}{\tilde{\vh}}
\newcommand{\auxParamTwo}{\tilde{\vJ}}
\newcommand{\auxCovariance}{\tilde{\vV}}

\newcommand{\meanVariationalParams}{\boldsymbol{\mu}}
\newcommand{\meanVariationalParamsTwo}{\boldsymbol{\Psi}}
\newcommand{\E}{\mathbb{E}}

\newcommand{\KL}[2]{\mathbb{D}_{\text{KL}}\!\left({#1}\lvert\rvert{#2}\right)}

\newcommand{\boldzero}{\boldsymbol{0}}

\DeclareMathOperator{\bigO}{\mathcal{O}}
\DeclareMathOperator{\cov}{cov}

\icmltitlerunning{Linear time GPs for neural data}

\begin{document}
\twocolumn[
    \icmltitle{Linear Time GPs for Inferring Latent Trajectories from Neural Spike Trains}

    \icmlsetsymbol{equal}{*}

    \begin{icmlauthorlist}
    \icmlauthor{Matthew Dowling}{sbu}
    \icmlauthor{Yuan Zhao}{nih}
    \icmlauthor{Il Memming Park}{ccu}
    \end{icmlauthorlist}

    \icmlaffiliation{sbu}{Stony Brook University, New York, USA}
    \icmlaffiliation{nih}{National Institute of Mental Health, USA}
    \icmlaffiliation{ccu}{Champalimaud Research, Champalimaud Foundation, Portugal}

    \icmlcorrespondingauthor{Matthew Dowling}{matthew.dowling@stonybrook.edu}
    \icmlkeywords{Machine Learning, ICML}

    \vskip 0.3in
]
\printAffiliationsAndNotice{}
\begin{abstract}
Latent Gaussian process (GP) models are widely used in neuroscience to uncover hidden state evolutions from sequential observations, mainly in neural activity recordings.
While latent GP models provide a principled and powerful solution in theory, the intractable posterior in non-conjugate settings necessitates approximate inference schemes, which may lack scalability.
In this work, we propose \CVIHMGP, a general inference framework for latent GP models leveraging \HidaMatern kernels and conjugate computation variational inference (CVI).
With \CVIHMGP, we are able to perform variational inference of latent neural trajectories with linear time complexity for arbitrary likelihoods.
The reparameterization of stationary kernels using \HidaMatern GPs helps us connect the latent variable models that encode prior assumptions through dynamical systems to those that encode trajectory assumptions through GPs.
In contrast to previous work, we use bidirectional information filtering, leading to a more concise implementation.
Furthermore, we employ the \textit{Whittle} approximate likelihood to achieve highly efficient hyperparameter learning.
\end{abstract}

\section{Introduction}
Arguably the spatiotemporal structure of neural population activity implements neural computation.
Although it is not directly observable, recovery of the (effective) latent neural state evolution from recordings of neural activity is possible~\cite{Paninski2009,Kao2015b}, and is critical for advancing our understanding of neural computation.
Strong experimental evidence supporting the existence of low dimensional neural manifolds has fueled research into developing statistical models that can be used to infer the dynamics underlying neural computation~\cite{macke_plds_2011,Pfau2013,Archer2014f,Frigola2014,Pandarinath_2018}.
These methods usually fall under the header of latent variable models (LVMs) and posit that the observed neural activity can be sufficiently explained by linear or nonlinear mappings of latent dynamics~\cite{PeiYe2021NeuralLatents}.
Among those, a large class of LVMs employ Gaussian processes (GPs) to specify \emph{a priori} beliefs on the temporal structure of latent trajectories~\cite{yu_cunningham_gpfa_2009,vlgp_zhao_2017,Koyama2010,bayesian_gpfa_2021}.

The success and ubiquitous use of GPs is due in part to favorable properties such as universality, flexibility, and intuitive control over smoothness via time/length scale and differentiablility.
However, GP inference generally lacks scalability and non-Gaussian observations make the exact posterior intractable.
Though approximate methods like sparse GPs~\cite{titsias_sparse_variational_paper} can help, this comes at the price of accuracy and expressiveness.
Variational inference is widely used to enable computationally efficient approximations, however, a naive implementation still requires solving a large and dense linear system with time complexity of $\bigO(T^3)$ and space complexity $\bigO(T^2)$ for a sequence of length $T$.

In this work, we combine two recent developments in GP inference, the \HidaMatern (HM) kernels and conjugate computation variational inference (CVI).
The linear state space representation of GP through HM kernels~\cite{dowling_sokol_hida_matern_2021} allows for efficient latent trajectory inference via Kalman filtering/smoothing.
Utilizing natural gradients of the exponential family, CVI further reduces VI to a numerically elegant iterative optimization~\cite{khan_cvi_17}.
As a result, the \textbf{conjugate variational \HidaMatern GP (\CVIHMGP)} framework accelerates latent GP inference to linear time while maintaining flexibility of kernel choice and computationally efficient hyperparameter optimization. Furthermore, we introduce the Whittle (marginal) likelihood as an attractive approximation for GP hyperparameter learning.

Our contributions are the following: 
\textbf{(i)} We propose \CVIHMGP,  combining \HidaMatern GPs and CVI , as a tool for extracting latent trajectories from multivariate (neural) time series in linear time complexity.  
\textbf{(ii)} We show that the \textit{information} filter in tandem with CVI results in a more concise inference scheme than the mean/variance Kalman filter; an added bonus of using the information filter is that natural parameters returned from simultaneous forward/backward filters can be combined additively to give us the natural parameters of the posterior.
\textbf{(iii)} We show that the Whittle likelihood approximation is more sample efficient for hyperparameter optimization and has a favorable time complexity of $\bigO(LT\log T)$, compared to $\bigO(TL_S^3)$ of (the lower bound of) the marginal log-likelihood, where $L_S \geq L$.

\section{Background}
\subsection{GP models of latent trajectories}\label{section:latent_gp_models}
In this work, we are interested LVMs that define a linear/nonlinear mapping \eqref{eq:obs} between the latent state and observed variables, and impose assumptions on the temporal structure of latent state evolution through a GP prior \eqref{eq:gp},
\begin{align}
	&z_l(t) \sim \mathcal{GP}(m_l(t), k_l(t, t')) \,\, &\text{(latent processes)} \label{eq:gp} \\
	&\vy_t \mid \vz_t \sim P\left(\vy \mid g(\vz_t)\right) \,\,& \text{(observation model)} \label{eq:obs}
\end{align}
where $\vz_t = (z_1(t), \ldots, z_L(t))^\top$ is one of the $L$ unobserved latent processes modeled by a GP with mean and covariance functions $m_l$ and $k_l$ respectively, and $\vy_t \in \reals^N$ represents the observation at time $t$ that probabilistically depends on the temporal slice of all latent processes at time $t$, $\vz_t \in \reals^{\numLatents}$. Once the data has been observed, the goal of Bayesian inference is to find the posterior distribution of latent processes,
$
	p(\vz_{1:L} \mid \vy_{1:T})
$,
as well as the (hyper-)parameters of $m_l$, $k_l$ and $g$. 
Without loss of generality, we follow the standard practice of setting the mean function to be zero.

The linear and Gaussian assumption provides a convenient parameterization for the observation model (e.g. GPFA~\cite{yu_cunningham_gpfa_2009})
\begin{equation}\label{eq:lingauss}
	\vy_{t} = \vC \latents_t + \vb + \boldsymbol{\nu}_{t} 
\end{equation}
where $\vC$ is a readout matrix, $\vb$ is a bias, and $\boldsymbol{\nu}_t\sim\mathcal{N}(\boldzero, \vR)$.
The linear Gaussian assumption makes it natural to appeal to Expectation-Maximization (EM) for learning hyperparameters, since the complete data log-likelihood, and posterior can be evaluated in closed form~\cite{yu_cunningham_gpfa_2009, dempster_em_1977}.
However, evaluating these closed form expressions to compute the posterior still requires solving a large linear system of equations~\cite{Rasmussen2005}.

Despite the convenience of a tractable posterior, the linear Gaussian observation model is not always suitable for various types of observations, e.g. point processes. Many methods~\cite{macke_plds_2011,Adam_2016,vlgp_zhao_2017,Pandarinath_2018} thus relax the assumption to non-Gaussian conditional distributions, i.e.
\begin{equation}
    y_{n,t} \mid \latents_t \sim p\left( y_{n, t} \mid g(\latents_t) \right) \label{eq:poisson_lik}
\end{equation}
where $n$ indexes the observation dimension and $g$ is a generic function.
The price to pay for using a non-Gaussian likelihood is an intractable posterior, necessitating the use of approximate Bayesian methods. %
Variational inference methods combat this by choosing a tractable family of distributions to approximate the posterior, for instance, a factorized Gaussian $q(\vz) = \prod_{l=1}^L \mathcal{N}(\vm_l, \vP_l)$, so that the evidence lower bound (ELBO),
\begin{equation}\label{eq:vlgp_elbo}
    \mathcal{L} = \E_{q(\vz)} \log p(\vy_{1:T} \mid \vz_1,\ldots,\vz_L) - \KL{q(\vz)}{p(\vz)}
\end{equation}
is maximized with respect to the variational parameters $(\vm_l, \vP_l)_{l=1}^L$.
However, variational inference suffers from similar scalability issues as exact GP inference; computation of the KL divergence between Gaussian distributions with unstructured covariance matrices scales as $\mathcal{O}(T^3)$.

\subsection{\HidaMatern GPs: A state space view of GP}\label{section:hida_matern}

The state-space model (SSM) representation of stationary GPs, that are finitely differentiable in mean-square, has proven itself as a useful tool for reducing the time complexity of GP posterior inference~\cite{sarkka_hartikainen_ssms_2010,chang_vi_ssm_gp_2020,solin_infinite_horizon_gp}.
It is easy to construct state-space representations of GPs when their kernel can be written as a linear combination of \HidaMatern kernels; in that case, the GP in tandem with its mean square derivatives is Markovian, %
which allows for the use of fast inference routines~\cite{hida_gp_book, levy_canonical_gps}.
Furthermore, the linear combinations of \HidaMatern kernels can approximate the covariance function of \textit{any} stationary GP~\cite{dowling_sokol_hida_matern_2021}, making them arbitrarily expressive.

Consider a stationary GP with an $M$-th order \HidaMatern kernel,
\begin{align}
	\cov(z_t, z_{t+\tau}) &= k_{H,M}(\tau; \, \sigma^2, b, \rho) \nonumber \\
    &= \sigma^2 \cos(2\pi b\tau) k_{\text{\Matern}}(\tau; \, \rho, M+\tfrac{1}{2})
\end{align} 
where $k_{\text{\Matern}}(\tau; \, \rho, M+\tfrac{1}{2})$ is the \Matern kernel with length-scale $\rho$ and smoothness parameter $\nu=M+\tfrac{1}{2}$~\cite{Rasmussen2005}.  
Such a GP is $M$ times differentiable in the mean-squared sense~\cite{Jazwinski2007-yx}. Even though a mean square differentiable GP \textit{is not} always Markovian, the vector process $\latents_t^S=[z_t, z_t^{(1)}, \ldots, z_t^{(M)}]$ \textit{is} Markovian, where $z_s^{(i)}$ is the $i^\text{th}$ mean square derivative of $z_s$.
Since Gaussainity is preserved under linear operations, the mean square derivative of a GP is also a GP, and so it is important to be able to compute the multi-output covariance function between a GP and its mean square derivatives~\cite{alvarezComputationallyEfficientConvolved2011}.  Fortunately, computing the multi-output covariance between a GP and its mean square derivatives only requires computing appropriate derivatives of the kernel function as we have the relation that $\cov(z_t^{(i)}, z_{t+\tau}^{(j)}) = (-1)^{j} k^{(i+j)}(\tau)$ with $k^{(i+j)}(\tau)$, the $(i+j)^\text{th}$ derivative of $k(\tau)$ with respect to $\tau$.
Thus the joint distribution between any two time points in the vector process is
\begin{align}
    p(\latents_{t+\tau}^S, \latents_t^S) &= \mathcal{N}\left(\begin{bmatrix} \VecZero\\ \VecZero \end{bmatrix},\, \begin{bmatrix} \vK(0) & \vK(\tau)\\ \,\,\,\,\vK(\tau)^\top & \vK(0) \end{bmatrix} \right)
\end{align}
where $\left[ \HidaMatrixDiff(\tau) \right]_{ij} = (-1)^j k^{(i+j)}(\tau)$ is the covariance matrix between the GP and its mean square derivatives $\tau$ time units apart. Now, as a result of the Markov property, the joint distribution of a \HidaMatern GP can be factored as $p(\latents_1) \prod p(\latents_{t} \mid \latents_{t-1})$;
using the marginalization property of Gaussian distributions, we can explicitly describe the generative process underlying the GP as the following linear dynamical system (LDS)~\cite{dowling_sokol_hida_matern_2021},%
\begin{equation}
	\latents_{t+\tau}^S = \vA(\tau) \latents_t^S + \vQ(\tau), \quad
	\vQ(\tau) \sim \mathcal{N}(\boldzero, \vQ(\tau))
\end{equation}
where
\begin{align}
	&\HidaStateTransition(\tau) = \vK(\tau) \vK(0)^{-1}\\
	&\vQ(\tau) = \vK(0) - \vK(\tau) \vK(0)^{-1} \vK(\tau)^{\top}
\end{align}

\textbf{State space representation of latent GP models}\hspace{0.5em}The covariance of the vector process $\latents_{t}^S$ coincides exactly with the covariance of the multi-output GP defined by the kernel $\HidaMatrixDiff(\tau)$.  With the representation of stationary and finitely differentiable GPs through the state-space representation of the \HidaMatern GPs, we can rewrite the generative model defined by Eq.~\eqref{eq:gp} and Eq.~\eqref{eq:lingauss} as follows
\begin{align}
	\latents_{l, t+\tau}^S &= \vA_l(\tau) \latents_{l, t}^S + \HidaStateNoise_l(\tau)\label{eq::gen_model_gp}\\
	\vy_{t} &= \vC \vH \latents_{t}^S + \vb + \boldsymbol{\nu}_{t}\label{eq::gen_model_gauss}
\end{align}
where $\latents_t^S$ is formed by stacking $L$ vector processes, i.e. $\latents_t^S = [\vz_{1,t}^S, \ldots, \vz_{L, t}^S]$, $\vH$ is a $L \times L_S$ selector matrix that extracts $\vz_t$ from $\vz_t^S$, and $L_S = \sum\nolimits_{l=1}^L M_l$ is the dimension of the \textit{extended state space} -- the dimensionality required to represent all $L$ latent processes in addition to their mean square derivatives.
The LDS formulation alleviates unfavorable computational complexity, allows inference through the well known Kalman filter and smoothing algorithms, and allows posterior inference in linear time with respect to sequence length~\cite{anderson_moore_state_space_book,Sarkka2011-do}. In addition, framing the problem in this manner facilitates the use of tools and techniques from the vast literature of Gaussian linear dynamical systems.  Application of the Kalman filter and smoothing algorithms to infer the posterior admit time complexity of $\bigO(L_S^3T)$ but since $L_S \ll T$ the price is negligible; in the case of large $L_S$ an asymptotic version of the state-space model can be used to avoid expensive operations in the smoothing step~\cite{solin_infinite_horizon_gp}.
Thus, given a stationary GP kernel (component) with exactly $M$ derivatives, there exists an equivalent LDS GP formed by appending $M-1$ extra latent dimensions.

\subsection{Conjugate variational inference}\label{sec:CVI}
Adapting the state-space representation \eqref{eq::gen_model_gp}, we can rewrite the general form of observation model \eqref{eq:poisson_lik} into
\begin{align}
	\vy_t \mid \latents_t &\sim p\left(\vy_t \mid g\left(\vH \latents_t^{S}\right)\right)\label{eq::gen_model_poisson}
\end{align}
Unfortunately, any non-Gaussian likelihood prohibits the immediate use of Kalman filtering and smoothing, and obscures the path to computationally feasible inference.  

Recent works~\cite{wilkinson_spatiotemporal_vgp_2021,chang_vi_ssm_gp_2020} have demonstrated how conjugate computation variational inference~\cite{khan_cvi_17} can be used to exploit the SSM representation of GPs for linear time approximate inference.  In a nutshell, CVI takes advantage of the fact that, when the variational approximation and prior are in the same exponential family, one step of natural gradient ascent\footnote{In App.~\ref{app:section:cvi}, we provide necessary details about exponential family distributions and natural gradient descent} on the ELBO reduces to a conjugate Bayesian update.  
Supposing that $q(\vz \mid \naturalVariationalParams) \approx p(\vz \mid \vy)$ is the variational posterior with natural parameter $\naturalVariationalParams$ and mean parameter $\meanVariationalParams$, and $p(\vz \mid \naturalVariationalParams_0)$ is the prior with natural parameter $\naturalVariationalParams_0$, one natural gradient step on the ELBO with learning rate $\alpha$ is equivalent to
\begin{align}
    &q(\vz \mid \naturalVariationalParams_{k+1}) \propto 
	\underbrace{\exp\left(\auxNaturalVariationalParams_k^\top \vT(\vz)\right)}_{\propto p(\auxLikelihood_k \mid \vz)} \, p(\vz\mid\naturalVariationalParams_0)\label{eq::cvi_first_step}\\
    &\auxNaturalVariationalParams_{k+1} = (1-\alpha_k) \auxNaturalVariationalParams_k\nonumber \\
    &\qquad\quad+ \alpha_k \sum\nolimits_{t} \nabla_{\meanVariationalParams_{k+1}} \E_{q(z_t \mid \naturalVariationalParams_{k+1})} \log p(y_t \mid z_t)\label{eq::cvi_second_step}
\end{align}
where $\tilde{\naturalVariationalParams}$ are auxiliary variables that can be considered natural parameters of \textit{pseudo observations} $\tilde{\vy}$, and $p(\auxLikelihood_k \mid \vz)$ is the exponential family distribution that would have $p(\vz \mid \naturalVariationalParams_0)$ as its conjugate prior.  A principled initialization for CVI is to set $\naturalVariationalParams_1 = \naturalVariationalParams_0$ and $\alpha_1=1$ so that $\auxNaturalVariationalParams_1 = \sum \nabla_{\meanVariationalParams} \E_{p(\vz \mid \naturalVariationalParams_0)}\log p(y_t \mid z_t)$. More details on CVI are provided in App.~\ref{app:section:cvi}.

\section{\CVIHMGP for non-conjugate latent GP models}
Combining \HidaMatern kernels and CVI, we propose conjugate variational \HidaMatern (\CVIHMGP), an efficient learning approach of (non-conjugate) latent GP models.
\subsection{Posterior inference}\label{section:cvhm_point_process}
Now, we are ready to formulate a procedure for posterior inference and parameter learning when the generative model of the data is specified according to Eq.~\ref{eq::gen_model_gp} and~\ref{eq::gen_model_poisson}.
In order to take advantage of CVI, we will consider variational Gaussian approximations, i.e. $q(\vz) = \mathcal{N}(\vm, \vP)$, that can be represented in their natural parameterization as
\begin{equation}
	q(\vz) =\exp\left(\vz^\top \vJ \vz + \vh^\top \vz - \log Z\right)
\end{equation}
with $\vz=(z_1, \ldots, z_T)^\top$, normalization constant $Z$, and natural parameters $\vJ = -\tfrac{1}{2} \vP^{-1} $,  $\vh=\vJ \vm$. 
For conciseness, we will exchangeably use $\naturalVariationalParams$ and $(\vh, \vJ)$.  A GP prior will have natural parameters $\naturalVariationalParams_0 = (\boldzero, -\tfrac{1}{2} \vK_{TT}^{-1})$ where $\vK_{TT}$ is the Grammian evaluated over $T$ points.
Since the prior is a GP, the conjugate likelihood is also Gaussian and we can denote the pseudo natural parameters by $(\auxParamOne, \auxParamTwo)$.   These parameters can then be converted into Gaussian pseudo observations $\auxLikelihood = \auxCovariance \auxParamOne$ with mean $\vz$ and covariance $\auxCovariance=-\tfrac{1}{2}\auxParamTwo^{-1}$. 
The first step of a CVI iteration, as in Eq.~\eqref{eq::cvi_first_step}, can now be written in a more familiar form as the following GP regression problem 
\begin{align}
    q(\vz \mid \naturalVariationalParams_{k+1}) &\propto 
	\mathcal{N}(\auxLikelihood_k \mid \vz, \auxCovariance_k) \cdot \mathcal{N}(\vz \mid \boldzero, \vK_{TT})
	\label{eq:aux_update}
\end{align}
Then the pseudo natural parameters, $\auxParamOne$ and $\auxParamTwo$, can be updated using the mean parameter gradient as in Eq.~\eqref{eq::cvi_second_step}. This allows each CVI iteration to be done in $\bigO(T L_S^3)$ time, where the computational bottleneck is solving for the LDS posterior in Eq.~\eqref{eq:aux_update}  Refinement of the variational approximation through additional gradient steps proceeds by recomputing $(\auxLikelihood, \auxCovariance)$, solving the GP regression problem, then updating $(\auxParamOne, \auxParamTwo)$. We summarize this procedure in Alg.~\ref{app:alg:cvhm} in App.~\ref{app:cvhm_implementation_details}.

\begin{figure}[t]
    \centering
    \includegraphics[width=0.48\textwidth]{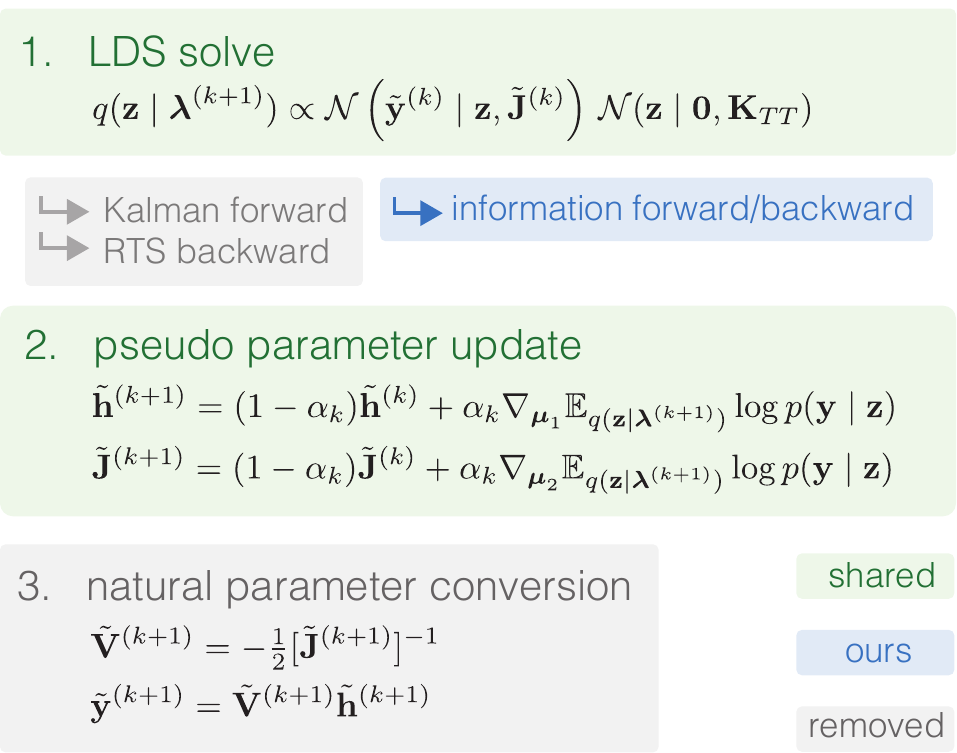}
    \caption{\textbf{Information filtering results in a more concise inference algorithm}\hspace{0.5em}By taking advantage of the pseudo observations being in the appropriate form for the information Kalman filter, we can avoid converting from the natural parameter representation.  Furthermore, information filtering forward/backward is easily parallelized and more numerically stable than filtering forward/smoothing backward.}
    \label{fig::algorithm_comparison}
\end{figure}

\begin{figure*}[ht]
    \centering
    \includegraphics[width=\textwidth]{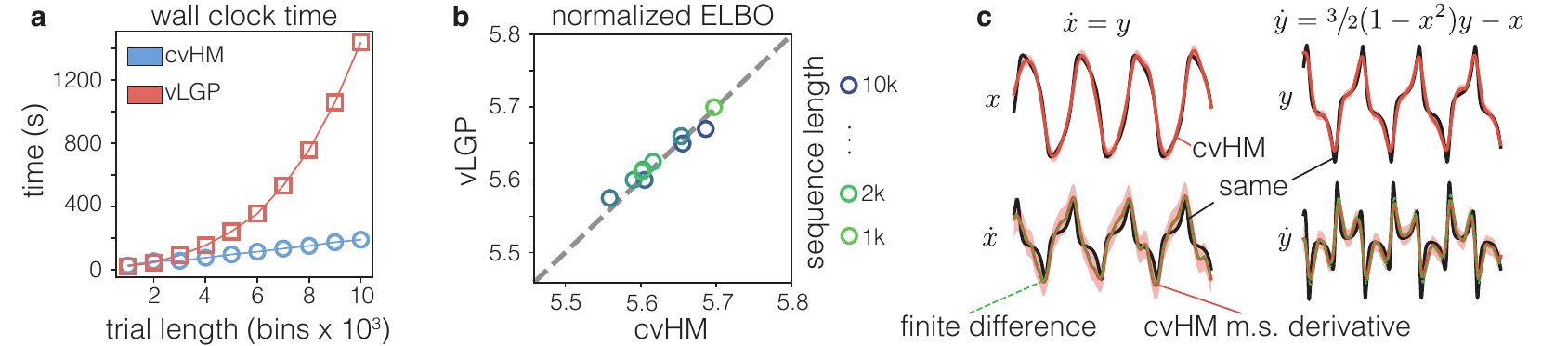}
    \caption{
    \textbf{(a)}
    \CVIHMGP inference scales better than vLGP. Sequence length varies from 1k to 10k in intervals of 1k. Hyperparameters of both methods are kept constant.
    \textbf{(b)}
    Normalized ELBO (nats/bin) comparison shows same quality inference.
    \textbf{(c)} Van der Pol oscillator experiment. (top) \CVIHMGP inference on the 2D latents.
    (bottom) Time derivative of $x$ and $y$, and corresponding mean square derivatives inferred by \CVIHMGP compared to finite difference of the inferred mean; GP inferred derivatives come at no additional cost and offer calibrated measure of uncertainty that may be useful in latent trajectory analysis to better understand neural computation.
    }
    \label{fig:vanderpol_meansq}
    \label{fig:vs_vlgp}
\end{figure*}

\textbf{Information Filtering} \hspace{0.5em}
Until now, we have glossed over algorithmic details of how the LDS posterior should be computed.
With the Gaussian pseudo observation, it is natural to use Kalman filter and RTS smoother to obtain the posterior mean and covariance. 
However, this requires the conversion from natural parameter space to mean-variance space (Fig.~1.3), $(\tilde{\vh}_t, \tilde{\vJ}_t) \mapsto (\auxLikelihood_t, \auxCovariance_t)$, every time after updating the pseudo natural parameters according to Eq.~\eqref{eq::cvi_second_step} (Fig.~1.2).
Not only does this conversion introduce additional computation each CVI iteration, it is liable to introduce numerical round off errors; in App.~\ref{app:ablation_studies}, we show that sidestepping these conversions in tandem with information filtering, results in improved inference at lower floating point precisions (results were similar using 64 bit floating point).

Fortunately, these conversions can be avoided if instead of computing the posterior through Kalman filtering/RTS smoothing we use the \textit{information} form of the Kalman filter~\cite{anderson_moore_state_space_book, kailath1980linear}.  In this situation we can think of posterior computation as a black box operation: the standard Kalman filter operates on $(\auxLikelihood_t, \auxCovariance_t)$, whereas the information form of the Kalman filter operates on $(\tilde{\vh}_t, \tilde{\vJ}_t)$, thus avoiding parameter conversions. How this change makes using CVI together with state-space GP priors more concise compared to other applications of CVI with state-space GPs in the literature, e.g.~\cite{chang_vi_ssm_gp_2020,wilkinson_spatiotemporal_vgp_2021}, is explained in Fig.~\ref{fig::algorithm_comparison}.

\textbf{Time-reversed dynamics and bidirectional filtering} \hspace{0.5em} Inference can be further accelerated by deviating from the practice of filtering forward/smoothing backward.  A favorable procedure, especially because we are working in the natural parameterization, is to take a message passing approach~\cite{bishop_pattern_recognition_2006}, and filter forward while filtering backward in parallel, then combine the statistics from each filter to compute the full posterior.  In order to perform backward filtering for an LDS, we require the backward representation of the generative process; for \HidaMatern GPs, the backward dynamics and backward state-noise are 
\begin{align}
    \vA^b(\tau) &= \vK(\tau)^{\top} \vK(0)^{-1}\\
    \vQ^b(\tau) &= \vK(0) - \vK(\tau)^\top \vK(0)^{-1} \vK(\tau)
\end{align}
which can be used to describe the generative process given by Eq.\ref{eq::gen_model_poisson} in backwards time,
\begin{align}
    \latents_{l, t}^S &= \vA_l^b(\tau) \latents_{l, t+\tau}^S + \HidaStateNoise_l^b(\tau)
\end{align}
Backward filtering returns the marginal posterior statistics of $p(\vz_t \mid \vy_{t:T})$, the filtering distribution at time $t$ given all data after that point (We provide more details about backward filtering in App.~\ref{app:backward_filtering}).  Thanks to the fact that the marginal prior, and backward/forward filtering approximate distributions are Gaussian (i.e. in the exponential family) if we factor the marginal posterior as
\begin{align}
    q(\vz_t \mid \vy_{1:T}) \propto \frac{\overbrace{q(\vz_t \mid \vy_{1:t})}^{\textcolor{orange}{\mathcal{N}(\vm_t^f, \vP_t^f)}} \overbrace{q(\vz_t \mid \vy_{t+1:T})}^{\textcolor{blue}{\mathcal{N}(\bar{\vm}_t^b, \bar{\vP}_t^b)}}}{\underbrace{p_{\priorHypers}(\vz_t)}_{\textcolor{gray}{\mathcal{N}(\boldzero, \vK(0))}}},
\end{align}
where $q(\vz_t \mid \vy_{t+1:T}) = \E_{q(\vz_{t+1}\mid \vy_{t+1:T})}[p_{\priorHypers}(\vz_{t+1}\mid\vz_t)]$ then, $q(\vz_t \mid \vy_{1:T})=\mathcal{N}(\vm_t, \vP_t)$, and the posterior marginal statistics can be directly read off
\begin{align}
    &\vP_t^{-1} = \textcolor{orange}{[\vP_t^f]^{-1}} + \textcolor{blue}{[\bar{\vP}_t^b]^{-1}} - \textcolor{gray}{\vK(0)^{-1}}\\
    &\vP_t^{-1} \vm_t = \textcolor{orange}{[\vP_t^f]^{-1} {\vm}_t^f} + \textcolor{blue}{[\bar{\vP}_t^b]^{-1} \bar{\vm}_t^b}
\end{align}
i.e. by adding the natural parameters recovered from the forward/backward filter, and subtracting natural parameters of the prior.  Thus, using forward/backward filters make computing the posterior as simple as combining the natural parameters returned from the forwards/backwards information filters. 
Thanks to stationary assumptions and linearity, the forward and backward filtering are mutually independent, and thus can be performed in parallel -- speeding up posterior inference by two-fold.  

\subsection{Learning GP Hyperparameters}
Standard practice for learning hyperparameters of the model is to use variational expectation maximization (VEM)~\cite{turner_problems_vEM}. For GP regression with a non-conjugate likelihood this would be a computational challenge;  every gradient step within the M-step requires recomputing the KL divergence term of the ELBO.   A useful result from~\cite{wilkinson_spatiotemporal_vgp_2021} is that the ELBO can be rewritten by using the GP regression form of the variational approximation to give
\begin{align}
	\mathcal{L}(\priorHypers) =& \sum_{t} \left[\E_{q(\latents_t)} \log \frac{p(\vy_t \mid \latents_t)}{p(\auxLikelihood_t \mid \latents_t)} + \log p_{\priorHypers}(\auxLikelihood_t \mid \auxLikelihood_{1:t-1})\right]
	\nonumber
\end{align}
where $p_{\priorHypers}(\auxLikelihood_t \mid \auxLikelihood_{1:t-1}) = \E_{\bar{q}_{\priorHypers}(\vz_{t})}[p(\auxLikelihood_t \mid \vz_{t})]$ and $\bar{q}_{\priorHypers}(\vz_{t}) = \E_{q(\vz_{t-1}\mid\auxLikelihood_{1:t-1})}[p_{\priorHypers}(\vz_t \mid \vz_{t-1})]$. Written this way, the ELBO can be evaluated in $\bigO(T L_S^3)$ time, due to the Markov structure of the variational approximation. The log-marginal likelihood of the auxiliary observations can be computed using the \textit{predicted} values of the Kalman filter used to compute the variational approximation.  However, the cubic scaling with $L_S$ could be prohibitive for models with a high-dimensional extended state-space.
\newline\newline\textbf{Spectral Hyperparameter Optimization} \hspace{0.5em} Although we can evaluate the ELBO analytically, lets first rewrite it dropping terms independent of $\priorHypers$, so that
\begin{align}
    \mathcal{L}(\priorHypers) &= -\KL{q(\vz)}{p_{\priorHypers}(\vz)} = - \E_{q(\vz)} \left[\log p_{\priorHypers}(\vz) \right] \label{eq::kl_likelihood}
\end{align}
Now, we will consider a \textit{spectral} approximation\footnote{We give an introduction to the Whittle likelihood in App.~\ref{app:whittle}} of the log-marginal likelihood given via \textit{Whittle's likelihood approximation}~\cite{whittle_spectral_1951,beran_long_memory_statistics}.
Let $S_{\priorHypers}(\omega) = \mathcal{F}[k_{\priorHypers}(\tau)]$ be the power spectral density (PSD) of the prior GP\footnote{$\mathcal{F}[\cdot]$ is the Fourier transform; $S_{\priorHypers}(\omega)$ \& $k_{\priorHypers}(\tau)$ are Fourier duals by the Wiener-Khintchine theorem.}, and $Z(\omega) = \mathcal{F}[\vz]$.
Then, by plugging in Whittle's approximate likelihood we have
\begin{align}
    \log p_{\priorHypers}(\vz) &= -\tfrac{1}{2}\left(\vz^\top \vK_{TT}^{-1} \vz + \log|\vK_{TT}|\right) + C\\
    &\approx -\tfrac{1}{2}\sum_j \left(\log S_{\priorHypers}(\omega_j) + \frac{||Z(\omega_j)||^2}{S_{\priorHypers}(\omega_j)}\right) \label{eq::whittle_integral_psd}
\end{align} 
From here, if we recall that the Fourier transform is a linear transform (e.g. $\mathcal{F}[\vz] = \vF \vz$ where $\vF \in \reals^{T\times T}$ is the DFT matrix) and Gaussianity is preserved under linear transformations, the expectation can be evaluated so that
\begin{align}
    \mathcal{L}(\priorHypers) &\approx -\tfrac{1}{2}\sum_j \left( \log S_{\priorHypers}(\omega_j) + \frac{\E_{q(\vz)} || \vf_j^\top \vz ||^2}{S_{\priorHypers}(\omega_j)} \right)\\
    &= -\tfrac{1}{2}\sum_j \left( \log S_{\priorHypers}(\omega_j) + \frac{\E_{q(a_j)} \left[a_j^2\right]}{S_{\priorHypers}(\omega_j)} \right)\label{eq:whittle_elbo}
\end{align}
where $q(a_j) = \mathcal{N}(a_j\mid \vf_j^\top \vm, \vf_j^\top \vP \vf_j)$ and $\vf_j$ is the $j^{\text{th}}$ row of the DFT matrix, $\vF$.  Evaluating this bound has an initial cost of $\bigO(LT \log T)$ for the Fourier transform of $L$ latent processes, but each gradient step on the hyperparameters is only as costly as evaluating the summation in Eq.~\eqref{eq::whittle_integral_psd}.  
\begin{figure}[t]
    \centering
    \includegraphics[width=0.48\textwidth]{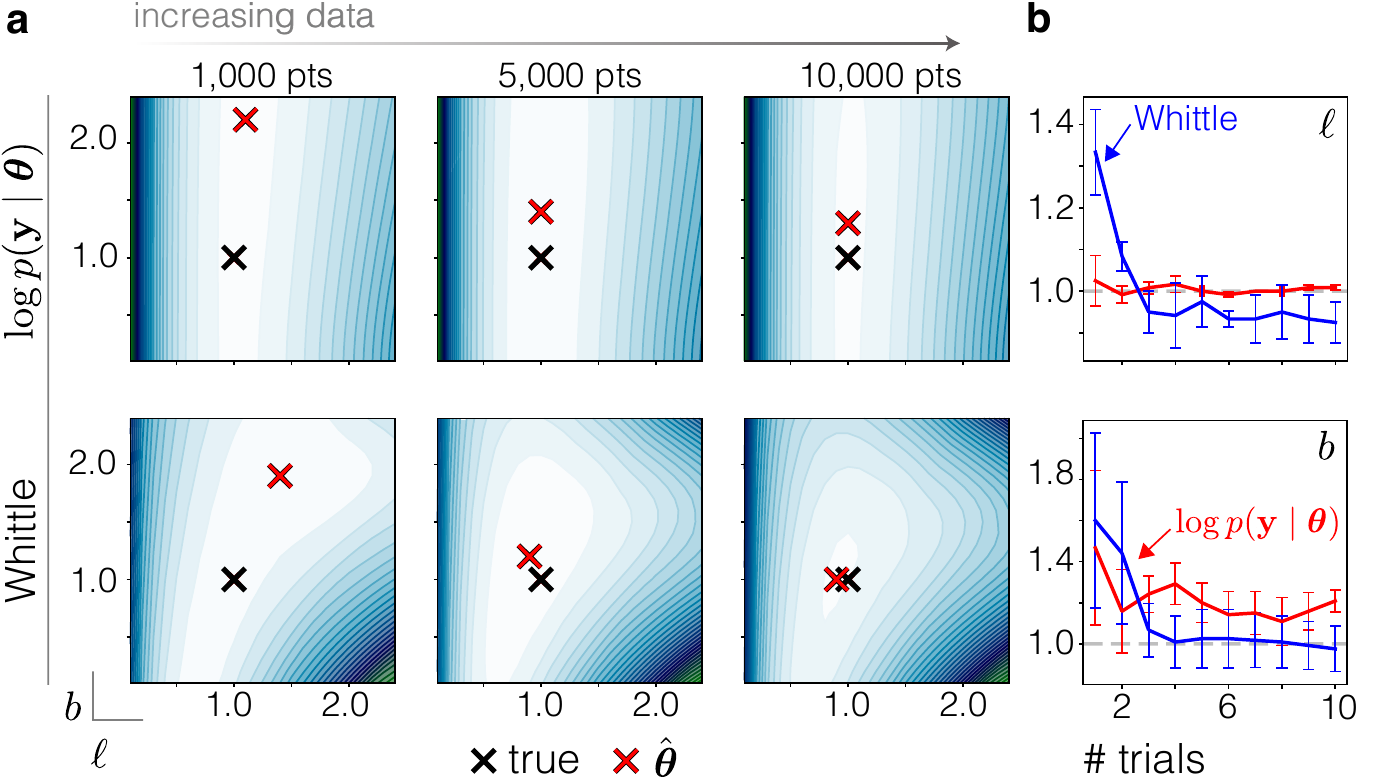}
    \caption{\textbf{(a)} We simulate a one-dimensional Gaussian LDS, where the latent variable are sampled from a GP with kernel $k_{H,1}(\tau; 1, 1, 1)$ then compare the log-marginal likelihood and the Whittle approximation as an objective for hyperparameter learning (from left to right 1k/5k/10k observations); $\hat{\priorHypers}$ denotes the parameter that maximizes objective.  The log-marginal likelihood does not seem to concentrate well on the frequency parameter. \textbf{(b)} Convergence of the log-marginal likelihood and Whittle approximation for hyperparameters averaged over 3 random seeds; the bias for the Whittle approximation is evident, but it also appears to have better convergence properties for $b$; we apply a Hann taper function before any Fourier transforms taken.}
    \label{fig::whittle_vs_lml}
\end{figure}
The optimal hyperparameters cannot be found in closed form, but the optimal PSD can, and may provide further intuition in regards to using the Whittle likelihood as an objective function.  The following proposition, which we derive in the Appendix, states that the PSD maximizing the ELBO at frequency $\omega_j$ is $\E_{q(a_j)} \left[a_j^2\right]$.
\begin{restatable}[Optimal $S_{\priorHypers}(\omega)$]{proposition}{optimalHyperparameters}\label{prop::whittle_squared}
    The function, $S^{\ast}_{\priorHypers}(\omega)$ maximizing the ELBO at frequencies $\omega_1, \ldots, \omega_{T/2}$, is given by
    \begin{align}
        S^{\ast}(\omega_j) &= \argmin_{S(\omega_j)} \,\, {\mathcal{L}}(S(\omega)) = \E_{q(a_j)} \left[a_j^2\right]
    \end{align}
\end{restatable}
This suggests that optimizing the Whittle likelihood attempts to bring the prior PSD closer to the expected periodogram of the posterior latent process.  In Fig.\ref{fig::whittle_vs_lml} we compare the Whittle likelihood to the log-marginal likelihood as an objective function.  The Whittle likelihood allows us to reduce hyperparameter optimization from $\bigO(L_S^3 T)$ to $\bigO(L T \log T)$; furthermore, the Whittle likelihood makes it possible to take advantage of methods/theory from signal processing in a probabilistic context.
\begin{figure*}[ht]
    \centering
    \includegraphics[width=\textwidth]{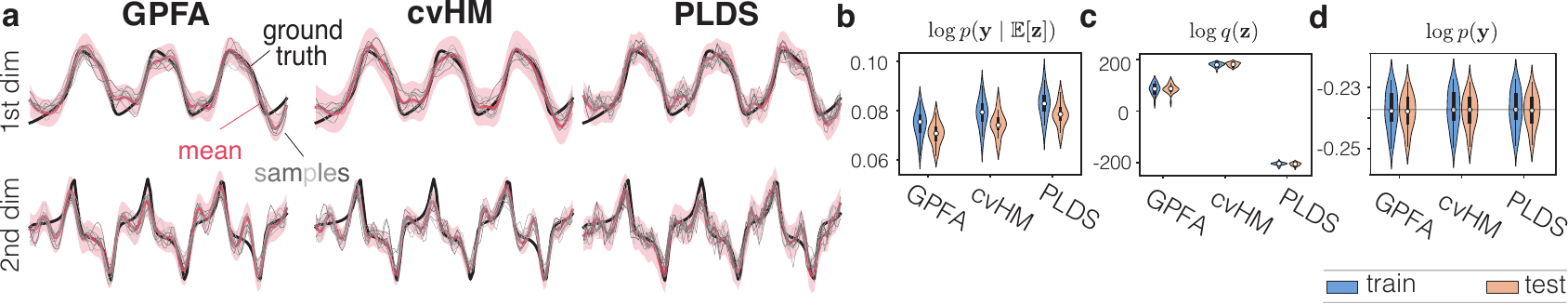}
    \caption{
    Van der Pol inference comparison of \CVIHMGP, GPFA, and PLDS.
    \textbf{(a)}
    Posterior means, credible intervals, and sample trajectories drawn from each posterior.
    Note that PLDS samples are much rougher, even though the posterior mean is smooth.
    \textbf{(b)}
    Reconstruction log-likelihood (bits per spike) based on the posterior mean trajectory~\cite{PeiYe2021NeuralLatents}.
    \textbf{(c)}
    Latent log-likelihood of the ground truth trajectory from the inferred posterior.
    \textbf{(d)}
    Marginal log-likelihood estimated via sampling.
    Gray line is centered at \CVIHMGP's median test marginal log-likelihood value.
    }
    \label{fig:vanderpol}
\end{figure*}
\section{Related work}\label{sec:related}
Unlike models e.g. \cite{lawrence05a} using GPs to define the functional relationship between latent and observed variables, the LVMs of our interest define a linear/nonlinear mapping between the latent state and observed variables, and use GP to impose \textit{a priori} temporal structure of latent trajectories;  P-GPLVM~\cite{wu2017} considers GP dynamics with a tuning curve function that is also modeled by GP.  Markovian linear autoregressive models, such as the Poisson Linear Dynamical System (PLDS), are extended to the nonlinear regime in~\cite{wang2005,Frigola2014,zhao2022a} by modeling the transition function with GP; unlike \CVIHMGP which aims to extract smooth latent trajectories, these methods in addition learn the underlying law that governs neural population dynamics.
\citet{chang_vi_ssm_gp_2020}, and~\citet{wilkinson_spatiotemporal_vgp_2021}, use CVI and the state-space representation of GPs for non-Gaussian observations, but in the context of spatio-temporal processes without necessarily considering a latent space.   

Although the state-space representation of \CVIHMGP with Poisson spiking is similar to PLDS~\cite{macke_plds_2011}, there are important conceptual differences:
PLDS explicitly specifies an LDS that governs the dynamics of the neural state, %
making it necessary to learn all parameters of the transition matrix.  Additionally, although the Laplace approximation used in PLDS is practically convenient, it negates theoretical guarantees of a monotonically increasing marginal likelihood provided by the EM algorithm. In contrast, the LDS in \CVIHMGP is determined by the small set of kernel hyperparameters, and is fixed to a degree of mean square differentiability.  However, encoding this structure in the prior comes at the cost of increased latent dimensionality and inferring latent variables which do not directly modulate the firing rate. Additionally, while our method applies to any model specifying prior beliefs through GPs finitely differentiable in mean square, such as those in~\cite{loper_leg_gp_2021, celerite_kernel, solin_periodic_kernel}, computing their state-space representation is not necessarily as straightforward as it is for a \HidaMatern GP.  In App.~\ref{app:markovian_gp_comparisons}, we examine the performance of \CVIHMGP on Markovian GP baselines such as those in~\cite{wilkinson2020state}.

\section{Experiments}

In this section, we put \CVIHMGP to the test on synthetic data and real neural recordings.
First, we verify on toy data that \CVIHMGP achieves the same performance as its latent GP model inference counterpart, vLGP~\cite{vlgp_zhao_2017}, but with linear time complexity.
Second, we compare \CVIHMGP with GPFA and PLDS on data generated according to dynamics of the Van der Pol system, a non-conservative oscillator with non-linear damping, to show its performance in the case of model mismatch.
Third, we apply \CVIHMGP on real neural recordings\footnote{\url{dandiarchive.org/dandiset/000130}, CC-BY-4.0} taken from a monkey performing a time interval reproduction task.
Finally, we demonstrate \CVIHMGP's potential impact on experimental design by showcasing its capability to handle long continuous neural recordings\footnote{\url{dandiarchive.org/dandiset/000129}, CC-BY-4.0}.

For synthetic data experiments, we use latent trajectories to generate spike trains from 150 neurons through a Poisson generalized linear model (GLM) with the canonical exponential as in Eq.~\eqref{eq:poisson_lik}.
The loading weights, $\vC$ and the bias, $\vb$, are drawn randomly.
We scale these parameters so that all neurons have a realistic baseline firing rate between 5 and 20 Hz.
For both real and synthetic neural data, we use 5~ms bins. %

\subsection{Linear time inference by \CVIHMGP}\label{section:experiment:vlgp}
To demonstrate that \CVIHMGP achieves the same performance as its counterpart with reduced computation time, we compare against vLGP which uses low-rank Cholesky decomposition to combat the time complexity.
In this example, we generated 1D latent trajectories from a \Matern 3/2 GP, with variance and length scale fixed to $1$ and $0.01$ respectively.
To illustrate the reduction in computational complexity, we vary the sequence length from $1,000$ up to $10,000$ bins, and run the experiments on the exact same computing setups to measure wall-clock time.
\CVIHMGP scales linearly (Fig.~\ref{fig:vs_vlgp}a) while achieving practically the same performance in terms of ELBO (Fig.~\ref{fig:vs_vlgp}b).

\subsection{Van der Pol oscillator}\label{section:experiment:van_der_pol}
We compare \CVIHMGP against PLDS~\cite{macke_plds_2011} and GPFA~\cite{yu_cunningham_gpfa_2009} to quantify how well it performs against models of the same class.
For this comparison, we simulate two-dimensional latent trajectories from the classic Van der Pol oscillator (Fig.~\ref{fig:vanderpol_meansq}c, top row).
Observed spike trains (Fig.~\ref{fig:app:trial_v_lengthscale} in App.~\ref{app:experimental_details}) were generated from the instantaneous latent states ($x(t)$ and $y(t)$).
We let all the methods optimize all the hyperparameters.

To quantify the goodness-of-fit for the inferred latent trajectories, we calculate the log-likelihood of the true (simulated) trajectories under the posterior inferred by each method, i.e. $\log q(\vz_{1:T})$.
The motivation for using this metric is that the reconstruction likelihood measure (Fig.~\ref{fig:vanderpol}b) used in the literature is designed to measure a deterministic firing rate prediction, forcing Bayesian approaches to disregard their posterior distributions.
This results in taking the mean latent trajectory of the posterior \textit{before} evaluating the log-likelihood of the observation.
In some cases, this may not give good insight into the quality of the inferred latent trajectories.
For instance, \CVIHMGP finds a higher quality variational posterior than GPFA~\footnote{\url{https://github.com/NeuralEnsemble/elephant}~\cite{elephant18}, BSD 3 license} and PLDS~\footnote{\url{github.com/lindermanlab/ssm}, MIT license} (Fig.~\ref{fig:vanderpol}a,c), while PLDS exhibits a higher reconstruction log-likelihood (Fig.~\ref{fig:vanderpol}a).
It is interesting that PLDS tends to trade off smoothness for a better fit to the firing rate since it has a worse posterior over latents. %
Unfortunately, the latent log-likelihood measure can only be evaluated in simulations where the true latent trajectory is known.
Since the time derivative of $x$ is equal to $y$ for the Van der Pol oscillator, we can compare the derivative process inferred by \CVIHMGP against its inferred trajectory for $y$.
In Fig.~\ref{fig:vanderpol_meansq}c we see that the mean square derivatives on average are similar to the analytic derivatives of the sampled trajectories, while providing a measure of calibrated uncertainty.  This illustrates how we can use \CVIHMGP to gain additional insights about aspects of latent trajectories that require knowing their time instantaneous derivatives, such as their velocity. 
\begin{figure}[ht]
    \centering
    \includegraphics[width=0.48\textwidth]{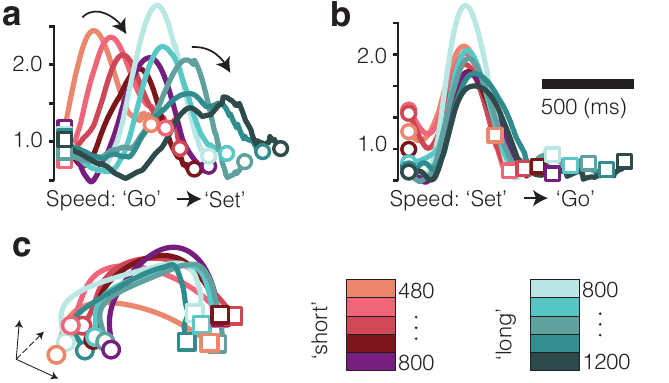}
    \caption{
    DMFC-RSG dataset `eye-right' condition.  
    \textbf{(a)} Speed of neural trajectories was easily inferred using the mean square derivatives; aligned to `Go'.  Results show that peak speed decreases within increasing intervals under each prior expectation (short vs.~long contexts, see~\cite{Sohn2019}).
    \textbf{(b)} Trajectory speeds aligned to `Set'.
    \textbf{(c)} Latent trajectories. 
    }
    \label{fig:realdata}
\end{figure}
\subsection{Electrophysiological Recordings} \label{section:experiment:time_interval}\label{section:experiment:long_trial}
\textbf{Time interval reproduction task}\hspace{0.5em}The utility of any LVM of neural dynamics is the ability to gain qualitative or quantitative understanding about neural computation by examining real data.  We use \CVIHMGP to examine the DMFC-RSG dataset~\cite{Sohn2019}. The DMFC-RSG dataset includes 54 neurons over 1289 trials recorded from dorso-medial frontal cortex (DMFC), and is known to exhibit low dimensional dynamics.  In this task a monkey is exposed to a timing interval demarcated by two visual cues, ``Ready'' and ``Set''.  Upon seeing the visual cue for ``Set'' the monkey waits an amount of time and signifies its prediction of the Ready-Set interval (marked ``Go'').  With \CVIHMGP we examine the inferred neural trajectories and their speeds using the mean square derivatives as shown in Fig.~\ref{fig:realdata}.  We find \CVIHMGP verifies the hypothesis that trajectories under the same prior should have speeds that decrease with increasing prediction interval.  To validate our results, we also ran GPFA and PLDS on this dataset and found the reconstruction log-likelihood evaluated on the training/test set of both methods to be similar (see Supplement).  Again, the utility of \CVIHMGP is apparent since to estimate the speed of these trajectories with PLDS and GPFA required using finite difference schemes which do not provide a calibrated measure of uncertainty.
\begin{figure}[ht]
    \centering
    \includegraphics[width=0.48\textwidth]{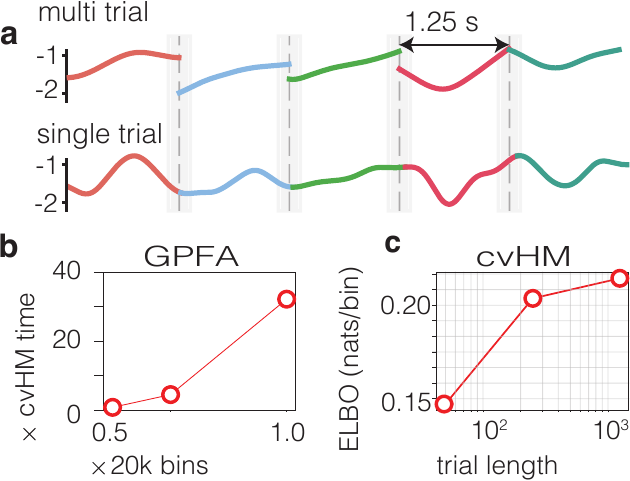}
    \caption{
    	\textbf{(a)} Staggered trajectories of a single latent dimension.
    	\textbf{(b)} GPFA inference computation time normalized by \CVIHMGP inference computation time; for long trials, GPFA becomes infeasible quickly.
    	\textbf{(c)} Additionally, decreasing trial length affects the quality of inference as measured by the ELBO; in addition to discontinuities, creating artificial trials affects hyperparameter adversely.
    }
    \label{fig:long_trial}
\end{figure}
\newline\newline\textbf{Long trial experiment}\hspace{0.5em}Many neuroscientific experiments have long trials or no inherent trial structure~\cite{ODoherty2017}.
To apply latent GP models, typically one has to split the recording into segments for the sake of computational cost.
Practical implementations split trials into even shorter segments for fast EM iterations (e.g. the GPFA implementation of \cite{elephant18}\footnote{https://github.com/NeuralEnsemble/elephant}).
These compromises have drawbacks on inference and hyperparameter tuning.
Fortunately, the linear time complexity of \CVIHMGP makes it feasible to analyze the continuous recordings in a reasonable amount of time.
In this experiment, we demonstrate this and show the effect of trial splitting using the MC-RTT dataset~\cite{ODoherty2017}.
This dataset consists of a 15 minute continuous recording from 130 neurons in motor cortex during a reaching task.
We use a spike train of 100$s$ (20,000 bins) as a single trial, and split it further into 50$s$ and 25$s$ trials. We fit \CVIHMGP to these 3 sets separately.

Figure~\ref{fig:long_trial}a shows how trial splitting affects the quality of inference.
Firstly, the inferred trajectories are staggered at the artificial trial boundaries.
Secondly, the inference deteriorates with split (Fig.~\ref{fig:long_trial}c).
Thirdly, the optimal length scale tends to become small as the trials get shorter (see Fig.~\ref{fig:app:length_scale_optimization} in the Appendix).
To show how inference can quickly become infeasible when using similar GP methods, we compare against GPFA.
In order to elucidate the benefit of the state-space approach we normalize sequence length against a 20,000 bin long segment, and wall clock time against \CVIHMGP's wall-clock time to infer the posterior of the 20,000 bin long segment (Fig.~\ref{fig:long_trial}b).

\section{Discussion}
In this work, we propose \CVIHMGP, a latent GP model learning approach combining the recent \HidaMatern framework and CVI.
We showed that \CVIHMGP provides competitive inference with favorable linear time complexity for non-conjugate observations.
Its high computational efficiency eliminates compromises such as the need to split long trials or use large time bins for practical analyses of long neural recordings, and opens a door to flexible experimental design.
Moreover, \CVIHMGP provides posterior beliefs about the mean square derivatives of the latent processes as a free lunch, which exhibits potentials in providing further insights for scientific questions. 
Furthermore, we introduced the Whittle likelihood as an alternative objective for hyperparameter learning of stationary GP models; in spite of bias, the Whittle likelihood approximation could accelerate GP inference as well as in other areas of machine learning. 

The GP prior, as we stated early, is essentially a linear dynamical system, so that \CVIHMGP, as well as other latent GP models, would mismatch the underlying dynamics if the ground truth is nonlinear. Nonetheless, this does not imply that the inferred trajectories come from a linear DS. 
Meanwhile, \HM theory restricts \CVIHMGP to stationary kernels, and could require high numerical precision if high order of derivatives are needed. For practicability, we have found it better to approximate smooth kernels with mixture of low-order \HM kernels to avoid ill numerical conditioning.

In future work, we aim to extend \CVIHMGP to handle autoregressive observations in order to capture salient details such as neuron refractory periods~\cite{Pillow2008,vlgp_zhao_2017}, to utilize nontrivial  observation models for nonlinear population coding (e.g. hippocampal CA1 place cells~\cite{wu2017}), and to account for control inputs in the dynamical system point of view. Moreover, we could employ the sparse GPs~\cite{sparse_ssm_gp_wilkinson_21, sparse_ssm_gp_adam_21} to further accelerate \CVIHMGP and the state-space representation and iterative nature of CVI may shed on light on online inference.

\section*{Acknowledgements}
MD and IP were supported by an NSF CAREER Award (IIS-1845836) and NIH RF1DA056404. YZ was supported in part by the National Institute of Mental Health Intramural Research Program (ZIC-MH002968). We thank the anonymous reviewers for their helpful feedback, comments, and motivating insightful experiments.

\newpage
\appendix
\onecolumn
\section{The Whittle likelihood}\label{app:whittle}
In the proposed inference procedure, the Whittle likelihood is used to reduce the computational complexity of hyperparameter optimization.  The Whittle likelihood has been used since its conception \cite{whittle_spectral_1951} to ameliorate the computational difficulties of evaluating the log-marginal likelihood of a Gaussian process; although this is a commonly used approximation in the stochastic processes literature~\cite{debiased_whittle_2019,reconciling_whittle,abstract_inference}, to the best of our knowledge this is the first time the Whittle likelihood has been used in the context of machine learning and for approximate Bayesian inference with non-Gaussian observations.

For the sake of accessibility, we will walk through a derivation of the Whittle likelihood; any of the aforementioned references can be consulted for further details.  Imagine we regularly sample a GP, $z(t) \sim \mathcal{GP}(0, k(\tau))$, at $T$ time points and collect these observations into the vector $\vz_{1:T} = \begin{bmatrix}z_1 & \ldots & z_T \end{bmatrix}$. Then, the log-likelihood of this particular sample is
\begin{align*}
    \log p(\vz_{1:T}) &= -\tfrac{1}{2}\left(\vz_{1:T}^\top \vK_{TT}^{-1} \vz_{1:T} + \log|\vK_{TT}| + T\log 2\pi\right)
\end{align*}

Unfortunately, evaluating $\vK_{TT}^{-1} \vz_{1:T}$ and $\log |\vK_{TT}|$ when $\vK_{TT}$ is dense scales on the order of $\bigO(T^3)$.  However, because the observations were regularly sampled and the kernel is stationary, $\vK_{TT}$ will be a Toeplitz matrix (constant along all diagonals).  The only necessary result to derive the Whittle likelihood, is a result from~\cite{grenander1958toeplitz} which states that
asymptotically, $\vK_{TT}$ can be decomposed as
\begin{align}
    \vK_{TT} \approx \vF^{H} \vD \vF
\end{align}
where $\vF$ is the DFT matrix, $\vD_{ii} = 1 / S_{\priorHypers}(\omega_i)$, and $H$ is the conjugate transpose. Using the fact that $\vF \vz_{1:T}$ is the DFT transform of $\vz_{1:T}$, we can plug in these expressions to arrive at the Whittle likelihood
\begin{align*}
    \log p(\vz_{1:T}) &\approx -\tfrac{1}{2}\sum_j \left(\log S_{\priorHypers}(\omega_j) + \frac{||Z(\omega_j)||^2}{S_{\priorHypers}(\omega_j)}\right)
\end{align*}
where $\omega_j = (2\pi j) / (\Delta T)$ with $\Delta$ the sampling interval.  The Whittle approximation is a biased estimate of the log-marginal likelihood, however, there exist improvements over the original approximation meant to reduce the approximation bias~\cite{debiased_whittle_2019}. While this bias is unfavorable, the reduction in time complexity of hyperparameter optimization from $\bigO(L_S^3 T)$ down to $\bigO(T \log T)$ is substantial.  Additionally, this allows tools/techniques from the signal processing literature for spectral estimation to be used in the context of probabilistic inference~\cite{reconciling_whittle}. In Fig.~\ref{fig::whittle_vs_lml} while effects of the bias are evident (i.e. estimation of $l$), the Whittle likelihood is substantially less invariant to changes in $b$; the elliptical landscape of the log-marginal likelihood, as is known for GPs~\cite{wu2017,Rasmussen2005}, complicates optimization.

\subsection{Optimal PSD}\label{app:proof:whittle_optimal_psd}
\optimalHyperparameters*
\textbf{Proof}:  \hspace{0.5em} Since terms couple additively across frequencies, we can just concern ourselves with finding
\begin{align}
    S^{\ast}(\omega_j) &= \argmax \, -\tfrac{1}{2} \left[ \log S_{\priorHypers}(\omega_j) + \frac{\E_{q(a_j)}[a_j^2]}{S_{\priorHypers}(\omega_j)} \right]
\end{align}
Now, we can regard $S_{\priorHypers}(\omega_j)$ as an ordinary variable -- setting the usual derivative to 0, we have that
\begin{align}
    &\frac{1}{S^{\ast}(\omega_j)} - \frac{\E_{q(a_j)}[a_j^2]}{S^{\ast}(\omega_j)^2} = 0\\
    \implies \,\, &S^{\ast}(\omega_j) = \E_{q(a_j)}[a_j^2]
\end{align}
so that the optimal PSD at each $\omega_j$ is just the expected magnitude of the squared DFT.
\section{Conjugate computation variational inference (CVI)}\label{app:section:cvi}
For completeness, we give a brief review of conjugate computation variational inference (CVI), for further details consult~\cite{khan_cvi_17,khan_lin_fast_simple_cvi_18}. CVI is applicable when we have a hierarchical Bayesian graphical model
\begin{align}
    \vz &\sim p_{\priorHypers}(\vz)\\
    \vy \mid \vz &\sim p(\vy \mid \vz)
\end{align}
and we want to find a variational approximation to the posterior, $q(\vz) \approx p(\vz \mid \vy)$.  If $q(\vz)$ and $p_{\priorHypers}(\vz)$ are chosen to be in the same exponential family of distributions~\cite{Wainwright2008}, then $q(\vz)$, can be factored as follows
\begin{equation}\label{eq:expfam}
	q(\vz) = h(\vz) \exp(\naturalVariationalParams^\top \vT(\vz) - A(\naturalVariationalParams))
\end{equation}
where $h(\vz)$ is the base measure, $\naturalVariationalParams$ is the natural parameter, $\vT(\vz)$ is the sufficient statistic, and $A(\naturalVariationalParams)$ is the log-partition function~\cite{khan_cvi_17}.  Parameters of the variational approximation are easily found through gradient ascent on the ELBO, i.e.
\begin{align}
    \naturalVariationalParams \leftarrow \naturalVariationalParams + \alpha \nabla_{\naturalVariationalParams} \mathcal{L}(\naturalVariationalParams)
\end{align}
However, the computation of the KL term can present a significant computational challenge -- scaling on the order of $\bigO(T^3L^3)$ for Gaussian distributions with dense and unstructured covariance matrices. Furthermore, gradients of the ELBO may not provide a suitable ascent direction in the space of probability distributions~\cite{hensman_natural_gradients_in_practice_2018}.  Alternatively, the natural gradient can be used to take advantage of the information geometry of exponential family distributions~\cite{amari_ngd_98}.  The natural gradients are given by $\vF(\naturalVariationalParams)^{-1} \nabla_{\naturalVariationalParams} \mathcal{L}$
where $\vF(\naturalVariationalParams) = \E_{q(\vz\mid\naturalVariationalParams)}[ {\nabla_{\naturalVariationalParams}^2} \log q(\vz)] = \nabla_{\naturalVariationalParams}^2 A(\naturalVariationalParams)$ is the Fisher information matrix.
The inverse Fisher information matrix prohibits naively employing natural gradient descent for large $T$.
However, the Fisher information matrix can be entirely avoided by considering another parameterization of the variational distribution in terms of its \emph{mean parameters}, defined as the expectation of the sufficient statistic\footnote{
The natural parameters should be in the \textit{minimal} exponential family form; here minimal is a technical requirement on the linear independence (always achievable) of sufficient statistics so that there exists a one-to-one mapping from the natural to the mean parameterization.
}
\begin{align}
	\meanVariationalParams(\naturalVariationalParams) &= \E_{q(\vz \mid \naturalVariationalParams)}[\vT(\vz)] = \nabla_{\naturalVariationalParams} A(\naturalVariationalParams)
\end{align}
where the natural parameters are also a function of the mean parameters\footnote{
The log-partition function in the natural parameters and the negative entropy in the mean parameters form a dual~\cite{Wainwright2008}.
}, i.e., we can write $\naturalVariationalParams(\meanVariationalParams)$.  By the chain rule, natural gradient (w.r.t. $\naturalVariationalParams$) of ELBO can be simply written as the gradient w.r.t. the corresponding mean parameters:
\begin{equation}
	\vF(\naturalVariationalParams)^{-1} \nabla_{\naturalVariationalParams} \mathcal{L} = \vF(\naturalVariationalParams)^{-1} \left( \nabla_{\naturalVariationalParams} \meanVariationalParams \right)\nabla_{\meanVariationalParams} \mathcal{L} = \nabla_{\meanVariationalParams} \mathcal{L}.
\end{equation}
Therefore gradient ascent on the ELBO can be done through updates without Fisher information matrix,
\begin{equation}
	\naturalVariationalParams_{k+1} = \naturalVariationalParams_k + \alpha_k \vF(\naturalVariationalParams_k)^{-1} \nabla_{\naturalVariationalParams_k}\mathcal{L} = \naturalVariationalParams_k + \alpha_k \nabla_{\meanVariationalParams_k}  \mathcal{L}.
\end{equation}
where $\alpha_k > 0$ are step sizes. Using these parameterizations,~\cite{khan_cvi_17} showed the Kullback-Leibler (KL) divergence between the prior and posterior simplifies as $\nabla_{\meanVariationalParams} \KL{q(\vz)}{p(\vz)} = \naturalVariationalParams_0 - \naturalVariationalParams$, where $\naturalVariationalParams_0$ are the natural parameters of the prior, $p(\vz)$.
Hence, the natural gradient ascent steps become
\begin{align}
    \naturalVariationalParams_{k+1} = \naturalVariationalParams_k + \alpha_k (\naturalVariationalParams_0 - \naturalVariationalParams_k) + \alpha_k \sum_t \nabla_{\meanVariationalParams_k} \E_{q(z_t)} \log p(y_t \mid z_t).
    \label{eq:natural_gradient_update}
\end{align}
Iterative updates can be transformed into the following two step procedure \cite{khan_cvi_17} using an auxiliary variable $\auxNaturalVariationalParams$ such that
\begin{align}
    \naturalVariationalParams_{k+1} &= \auxNaturalVariationalParams_k + \naturalVariationalParams_0 \label{eq:conjugate}\\
    \auxNaturalVariationalParams_{k+1} &= (1-\alpha_k) \auxNaturalVariationalParams_k + \alpha_k \sum_t \nabla_{\meanVariationalParams_k} \E_{q(z_t)} \log p(y_t \mid z_t) \label{eq:step2}
\end{align}
where care should be taken to note the dependence between $\meanVariationalParams_k$ and $\naturalVariationalParams_k$.
Note that \eqref{eq:conjugate} resembles a Bayesian posterior calculation with a conjugate prior, while \eqref{eq:step2} updates the natural parameters $\auxNaturalVariationalParams_k$ of the (approximate) likelihood.
This would suggest that the first step can be thought of as a Bayesian posterior calculation, where $\auxNaturalVariationalParams$ are associated with \textit{pseudo-observations} $\auxLikelihood$ conjugate to the prior.

\section{Probabilistic filtering for posterior inference}
\subsection{Backward filtering}\label{app:backward_filtering}
Consider the following LDS modeled forward in time,
\begin{align}
    \vz_{t+1} &= \vA \vz_t + \boldsymbol{\epsilon}_t\\
    \vy_t &= \vC \vz_t + \boldsymbol{\eta}_t
\end{align}
In our context, this model encodes our a priori belief about the latent dynamics.  Associated with the SSM that runs forward in time, is an SSM that runs backward in time 
\begin{align}
    \vz_t &= \vA^b \vz_{t+1} + \boldsymbol{\epsilon}^b_t\\
    \vy_t &= \vC \vz_{t} + \boldsymbol{\eta}_t
\end{align}
where $\boldsymbol{\epsilon}^b_t \sim \mathcal{N}(\boldzero, \vQ^b(\tau))$ is the backward state-noise and $\vA^b$ is the backward-time dynamics.  In general the backwards dynamics and state-noise satisfy the following equations~\cite{kailath1980linear}
\begin{align}
    \vA^b_{t+1} &= \vS_t \vA^{\top} \vS_t^{-1}\\
    \vQ^b_{t+1} &= \vS_t - \vA^b_{t+1} \vS_{t+1} \vA_{t+1}^{b\top}
\end{align}
where $\vS_t = \E_{p_{\priorHypers}(\vz_t, \vz_{t+1})}\left[\vz_t \vz_{t+1}^\top\right]$. When the prior dynamics is a \HidaMatern GP and all data are sampled at intervals of $\tau$, then $\vS_t = \vK(\tau)$ for all $t$ and the dynamics/state-noise are given by
\begin{align}
    \vA^b(\tau) &= \vK(\tau)^{\top} \vK(0)^{-1}\\
    \vQ^b(\tau) &= \vK(0) - \vK(\tau)^\top \vK(0)^{-1} \vK(\tau)
\end{align}
respectively.  Now, any filtering algorithm could be used in order to compute the backwards filtering distribution given by $p(\vz_t \mid \vy_{t:T})$.  For example, if we know $p(\vz_{t+1} \mid \vy_{t+1:T})$, then the predict and update steps of recursive inference are
\begin{align}
    p(\vz_{t} \mid \vy_{t+1:T}) &= \E_{p(\vz_{t+1} \mid \vy_{t+1:T})} \left[p_{\priorHypers}(\vz_{t} \mid \vz_{t+1})\right]\\
    p(\vz_t \mid \vy_{t:T}) &\propto p(\vy_t \mid \vz_t) \, p(\vz_t \mid \vy_{t+1:T})
\end{align}

\begin{algorithm}[tb]
    \caption{Backward information filter}
    \label{alg:example}
 \begin{algorithmic}[1]
    \STATE {\bfseries Input:} \, $\vy_{1:T}$, $\vV_{1:T}$, $\vA^b$, $\vQ^b$, $\vQ_0$
    \STATE initialization:
    \STATE \hspace{0.5em} $\vh_{T+1} \leftarrow \boldzero$
    \STATE \hspace{0.5em} $\vJ_{T+1} \leftarrow -\tfrac{1}{2} \vQ_0^{-1}$
    \STATE \hspace{0.5em} $\tilde{\vh}_{t} \leftarrow \vC^\top \vV_t^{-1} \vy_t \quad t=1,\ldots,T$
    \STATE \hspace{0.5em} $\tilde{\vJ}_{t} \leftarrow \vC^\top \vV_t^{-1} \vC \quad t=1,\ldots,T$
    \FOR{$t=T$ {\bfseries to} $1$}
        \STATE *prediction step*
        \STATE $\vL \leftarrow \vJ_{t+1} + \vA^{b\top} \vQ^{-b} \vA^b$
        \STATE $\bar{\vh}_t \leftarrow \vQ^{-b} \vA^b \vL^{-1} \vh_{t+1}$
        \STATE $\bar{\vJ}_t \leftarrow \vQ^{-b} - \vQ^{-b}\vA^{b} \vL^{-1} \vA^{b\top}\vQ^{-b}$
        \STATE *update step*
        \STATE $\vJ_t \leftarrow \bar{\vJ}_t + \tilde{\vJ}_t$
        \STATE $\vh_t \leftarrow \bar{\vJ}_t + \tilde{\vJ}_t$
    \ENDFOR
    \STATE {\bfseries Return:} \, $\vJ_{1:T}, \vh_{1:T}$
 \end{algorithmic}
 \end{algorithm}
\subsection{Combining statistics from forward/backward filters}
The dual filtering approach described in the main text is the preferred method for recovering statistics of the LDS posterior compared to more common filter forward/smooth backward algorithms -- especially in the context of variational inference where the natural parameterization is often more helpful.  More than that, dual filtering approach allows us to combine the natural parameters returned from both filters to easily compute posterior statistics.  

From the forward filter, we recover $q(\vz_t \mid \vy_{1:t})$ for all $t$, through the natural parameters $\textcolor{orange}{(\vh_t^f, \vJ_t^f)}$.  At the same time, from the backward filter, we recover $q(\vz_t \mid \vy_{t:T})$ through the natural parameters ${(\vh_t^b, \vJ_t^b)}$; however, we require the natural parameters, $\textcolor{blue}{(\bar{\vh}_t^b, \bar{\vJ}_t^b)}$,  of the backward predictive distribution, $q(\vz_t\mid \vy_{t+1:T})$, to compute the posterior.  Fortunately, they are just a byproduct of running the backward filter.
\begin{align}
    q(\vz_t \mid \vy_{1:T}) &\propto  q(\vy_{t+1:T}\mid \vz_t) q(\vz_t \mid \vy_{1:t})\\
    &\propto \frac{\overbrace{q(\vz_t \mid \vy_{1:t})}^{\textcolor{orange}{\mathcal{N}(\vm_t^f, \vP_t^f)}} \overbrace{q(\vz_t \mid \vy_{t+1:T})}^{\textcolor{blue}{\mathcal{N}(\bar{\vm}_t^b, \bar{\vP}_t^b)}}}{\underbrace{p_{\priorHypers}(\vz_t)}_{\textcolor{gray}{\mathcal{N}(\boldzero, \vK(0))}}},
\end{align}
where $q(\vz_t \mid \vy_{t+1:T}) = \E_{q(\vz_{t+1}\mid \vy_{t+1:T})}[p_{\priorHypers}(\vz_{t+1}\mid\vz_t)]$ then, $q(\vz_t \mid \vy_{1:T})=\mathcal{N}(\vm_t, \vP_t)$, and the posterior marginal statistics can be directly read off 
\begin{align}
    &\vP_t^{-1} = \textcolor{orange}{[\vP_t^f]^{-1}} + \textcolor{blue}{[\bar{\vP}_t^b]^{-1}} - \textcolor{gray}{\vK(0)^{-1}}\\
    &\vP_t^{-1} \vm_t = \textcolor{orange}{[\vP_t^f]^{-1} {\vm}_t^f} + \textcolor{blue}{[\bar{\vP}_t^b]^{-1} \bar{\vm}_t^b}
\end{align}
or, we can just substitute the natural parameter representation so that
\begin{align}
    \vJ_t &= \textcolor{orange}{\vJ_t^f} + \textcolor{blue}{\bar{\vJ}_t^b} - \textcolor{gray}{\vK(0)^{-1}}\\
    \vh_t &= \textcolor{orange}{\vh_t^f} + \textcolor{blue}{\bar{\vh}_t^b}
\end{align}
\subsection{Information filter}
Whereas, the traditional Kalman filtering algorithm uses recursive updates for the latent state mean/covariance, the information filtering algorithm uses recursive updates for the natural parameters of the latent state~\cite{anderson_moore_state_space_book}.  Recall, Kalman filter updates are given according to
\begin{align}
    \vP_t &= \bar{\vP}_t - \bar{\vP}_t \vC^\top (\vC \bar{\vP}_t \vC^\top + \vV_t)^{-1} \vC \bar{\vP}_t\\
    &= \left(\bar{\vP}_t^{-1} + \vC^\top \vV_t^{-1} \vC\right)^{-1}
\end{align}
so that,
\begin{align}
    \vP_t^{-1} &= \bar{\vP}_t^{-1} + \vC^\top \vV_t^{-1} \vC
\end{align}
Just as easily, we can plug in the expression for $\bar{\vP}_t = \vA\vP_{t-1}\vA^\top + \vQ$ -- invoking the Woodbury identity again reveals that
\begin{align}
    \bar{\vP}_t^{-1} &= \vQ - \vQ \vA^{\top} (\vA \vP_{t-1}^{-1} \vA^{\top} + \vQ^{-1})^{-1}  \vA \vQ
\end{align}
Illustrating that it could be advantageous to consider a recursion for $\vP_t^{-1}$ instead of $\vP_t$ -- especially if the latent dimensionality is significantly smaller than the number of neurons (which is often the case for neuroscience experiments).  More in depth treatment can be found in standard texts such as~\cite{anderson_moore_state_space_book, kailath1980linear}.

\section{Experimental details}\label{app:experimental_details}
\subsection{More comparison to the latent GP counterpart}
\begin{figure*}[htbp]
    \centering
    \includegraphics[width=\textwidth]{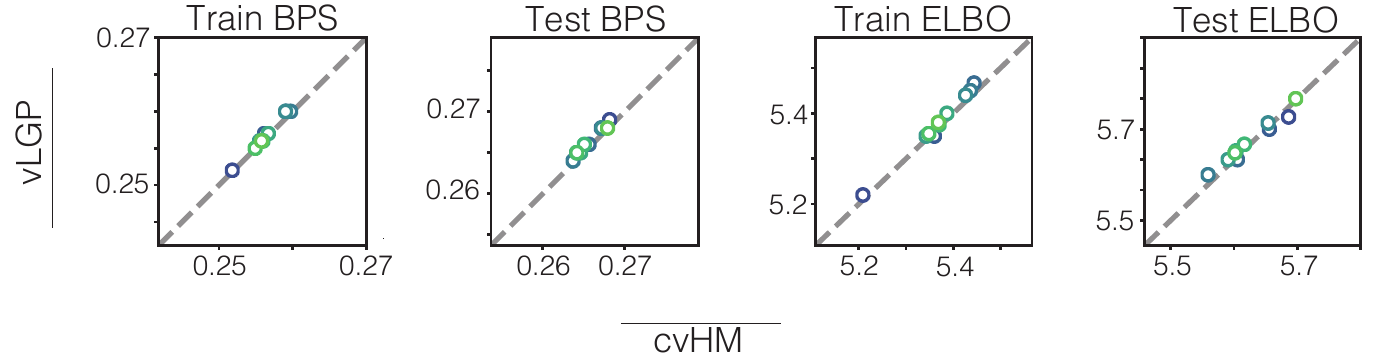}
    \caption{\CVIHMGP vs vLGP: BPS/normalized ELBO on train/test splits of the toy data used for verification. \CVIHMGP performs qualitatively the same as vLGP in terms of all metrics.}
    \label{fig:app:vlgp_cvhm_gof}
\end{figure*}

In the main text, we presented the normalized ELBO on a test set of the toy data generated from GP latents. Here we show BPS (bits-per-spike) and normalized ELBO measures in Fig.~\ref{fig:app:vlgp_cvhm_gof} to further verify that \CVIHMGP performs qualitatively the same as vLGP.

\subsection{Van der Pol oscillator predictions}
With the expressiveness of \HidaMatern kernels, we can predict (forecast) future latent states. For the Van der Pol oscillator experiment we wondered how well \CVIHMGP could predict future latent states in the absence of data. \begin{wrapfigure}[16]{r}{0.25\textwidth}
    \vspace{-10pt}
    \centering
    \includegraphics[width=0.25\textwidth]{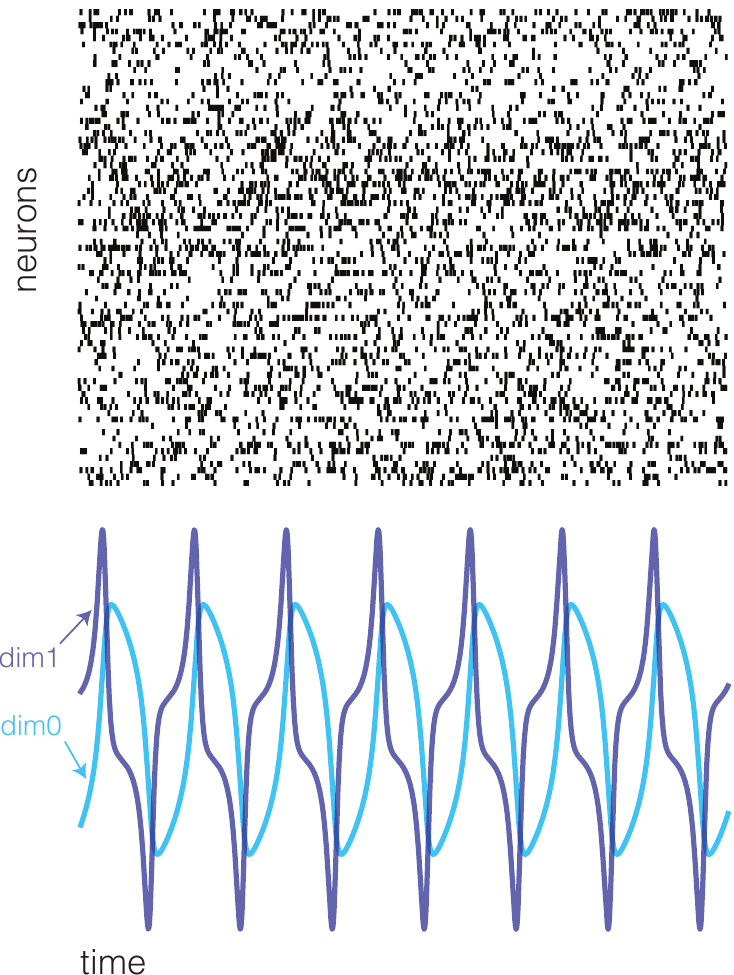}
    \caption{Observed spike counts and the `true' latent trajectories for the VDP system.}
    \label{fig:app:trial_v_lengthscale}
\end{wrapfigure}Since the Van der Pol system evolves according to nonlinear dynamics, a sufficiently expressive covariance function is needed per dimension in order to make accurate predictions. We can easily construct expressive \HidaMatern kernels through linear combination; we use 6, 2-ple \HidaMatern kernels for the $x$ dimension, and 30, 2-ple \HidaMatern kernels for the $y$ dimension.

\begin{figure*}[t]
	\centering
	\includegraphics[width=\textwidth]{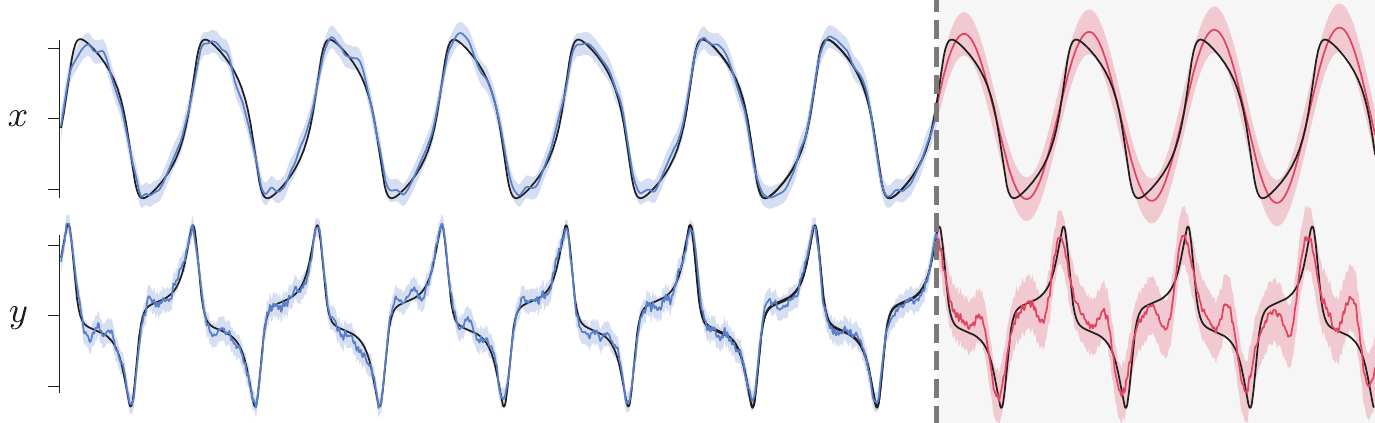}
	\caption{Van der Pol oscillator: \CVIHMGP predictions of future latent states of the.  
		Before dash line: approximate posterior mean in \textcolor{blue}{blue}, 
		After dash line: predictive mean in \textcolor{red}{red}
		Colored shade: the 95\% credible interval.
		Predictions of the $y$ dimension are able to capture salient features such as the sharp peak and asymmetric behavior about it. }
	\label{fig:app:vanderpol_predict}
\end{figure*}

The kernels over the $y$ dimension are initialized so that they cover a range of frequencies from 0 Hz to 70 Hz. In order to isolate how expressive a kernel we can create, we further initialize the loading matrix and bias to the true values.  During inference, we optimize all hyperparameters of all kernels. Fig.~\ref{fig:app:vanderpol_predict} shows that \CVIHMGP can perform prediction well at the cost of having a large expanded latent space.
\subsection{DMFC-RSG}
For the DMFC-RSG dataset, we analyzed all four conditions: `hand-left,' `hand-right,' `eye-left,' `eye-right' using \CVIHMGP, GPFA, and PLDS.  Using the different methods we examine time instantaneous speed of neural trajectories either aligned to `Set' or `Go'.  We fix the latent space to be three-dimensional for the purposes of visualization as well as qualitative metrics of performance.
\begin{figure*}[htbp]
    \centering
    \includegraphics[width=\textwidth]{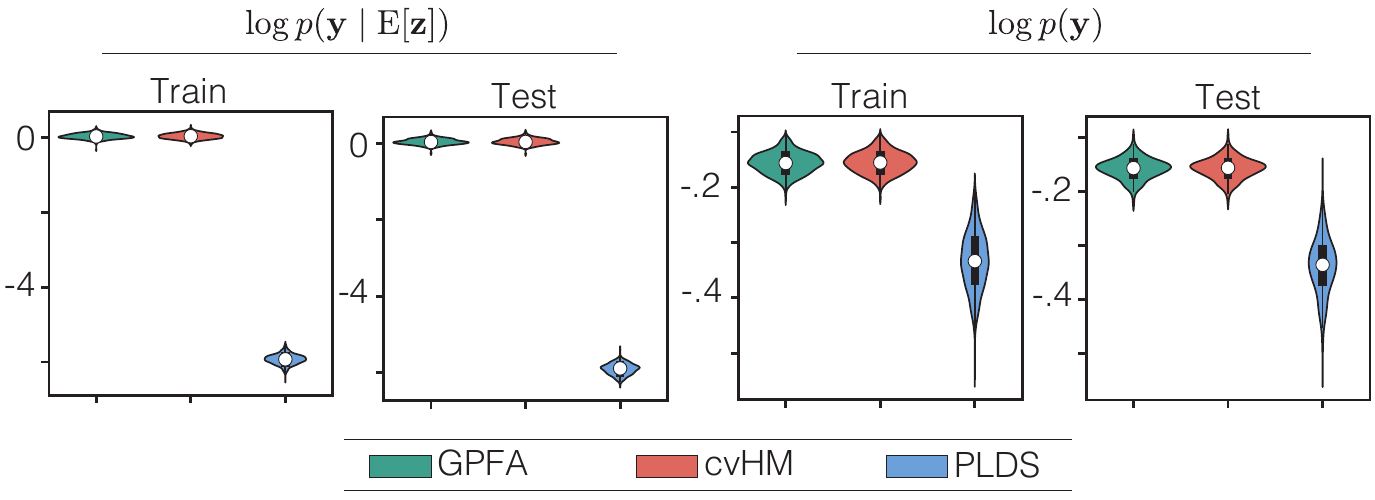}
    \caption{DMFC-RSG: BPS and log-marginal-likelihood for GPFA, \CVIHMGP, and PLDS per trial. Log-marginal-likelihood is calculated using a Monte-Carlo estimate with 100,000 samples from the posterior.  A soft rectification was used in order to transform the output of GPFA to fitting rate for log-marginal-likelihood.}
    \label{fig:app:dmfc_gof}
\end{figure*}
Fig.~\ref{fig:app:dmfc_gof} shows that in terms of BPS and log-marginal-likelihood, \CVIHMGP and GPFA perform similarly while PLDS lags behind. Be aware that the true dynamics in nonlinear and \CVIHMGP and GPFA have a larger effective dimensionality in SSM point of view.  In section~\ref{section:dmfc_figures}, the inferred speed of latent trajectories by all the three methods are plotted in Fig.~\ref{fig:app:dmfc_eye_left},~\ref{fig:app:dmfc_eye_right},~\ref{fig:app:dmfc_hand_left},and~\ref{fig:app:dmfc_hand_right}.

\subsection{MC-RTT}
To see how `trial splitting' affects the hyperparameter tuning, we examined the estimation of length scale using the MC-RTT dataset where we used a spike train of 20,000 bins but split it into trials of lengths 25, 50, 250, and 1250 bins.  In Fig.~\ref{fig:app:trial_v_lengthscale}, we see that the optimal length scale inferred for one of the latent dimensions decreases monotonically with increasing trial lengths.
\begin{wrapfigure}[16]{r}{0.2\textwidth}
    \centering
    \includegraphics[width=0.2\textwidth]{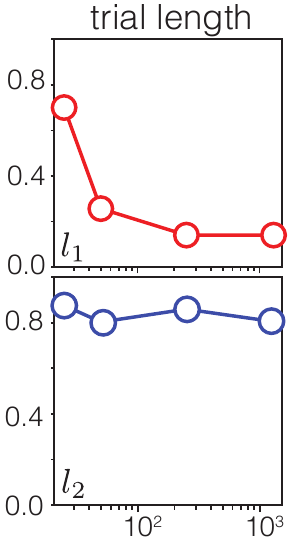}
    \caption{Inferred lengthscales of the two latent dimensions for MC-RTT as a function of trial length}
    \label{fig:app:trial_v_lengthscale}
\end{wrapfigure}
\section{\CVIHMGP implementation details}\label{app:cvhm_implementation_details}
\subsection{Initialization}
For \CVIHMGP, we initialize the readout matrix, $\vC$, using factor analysis and the bias, $\vb$, using the average firing rate.  Except for the first experiment to compare performance with vLGP, hyperparameters are optimized in variational EM style.  For optimization of the readout matrix, bias, and kernel hyperparameters we use PyTorch~\cite{pytorch_2019} in combination with SciPy~\cite{scipy_2020}.
\subsection{\CVIHMGP Algorithm}
We describe the exact \CVIHMGP learning algorithm that uses dual information filtering along with CVI for inference in Alg.~\ref{app:alg:cvhm}. There, $\vQ_{\infty}$ is the stationary covariance of the dynamical system defined by $(\vA, \vQ)$, and $\vh_0$ and $\vJ_0$ are the prior natural parameters; which in our case will be $\boldzero$ and $\vK_{TT}$ respectively.
\begin{algorithm}[tb]
    \caption{\CVIHMGP Algorithm}
    \label{app:alg:cvhm}
 \begin{algorithmic}[1]
    \STATE {\bfseries Input:} \, $\vy_{1:T}$, $\vV_{1:T}$, $\vA$, $\vQ$, $\vA^b$, $\vQ^b$, $\vQ_{\infty}$, $\vh_0$, $\vJ_0$
    \STATE initialization:
    \STATE \hspace{0.5em} *initialize the variational approximation to the prior*
    \STATE \hspace{0.5em} $\vh_{1:T} \leftarrow \vh_0$
    \STATE \hspace{0.5em} $\vJ_{1:T} \leftarrow \vJ_0$
    \STATE \hspace{0.5em} *initialize pseudo natural parameters*
    \STATE \hspace{0.5em} $\tilde{\vh}_{t} \leftarrow \nabla_{\meanVariationalParams_t^{(1)}} \E_{q(\vz_t)}[\log p(\vy_t \mid \vz_t)], \quad t=1,\ldots,T$
    \STATE \hspace{0.5em} $\tilde{\vJ}_{t} \leftarrow \nabla_{\meanVariationalParams_t^{(2)}} \E_{q(\vz_t)}[\log p(\vy_t \mid \vz_t)], \quad t=1,\ldots,T$
    \REPEAT
        \STATE *bidirectional filtering*
        \STATE $(\vh_{1:T}^f, \vJ_{1:T}^f) \leftarrow \text{FilterForward}(\tilde{\vh}_{1:T}, \tilde{\vJ}_{1:T}, \vA, \vQ, \vQ_{\infty})$
        \STATE $(\bar{\vh}_{1:T}^b, \bar{\vJ}_{1:T}^b) \leftarrow \text{FilterBackward}(\tilde{\vh}_{1:T}, \tilde{\vJ}_{1:T}, \vA^b, \vQ^b, \vQ_{\infty})$
        \STATE ${\vJ}_t \leftarrow \vJ_{1:T}^f + \bar{\vJ}_{1:T}^b - \vQ_{\infty}^{-1}$
        \STATE ${\vh}_t \leftarrow \vh_{1:T}^f + \bar{\vh}_{1:T}^b$
        \STATE *update pseudo natural parameters*
        \STATE $\tilde{\vh}_{t} \leftarrow (1-\alpha)\tilde{\vh}_t + \alpha\nabla_{\meanVariationalParams_t^{(1)}} \E_{q(\vz_t)}[\log p(\vy_t \mid \vz_t)], \quad t=1,\ldots,T$
        \STATE $\tilde{\vJ}_{t} \leftarrow (1-\alpha)\tilde{\vJ}_t + \alpha\nabla_{\meanVariationalParams_t^{(2)}} \E_{q(\vz_t)}[\log p(\vy_t \mid \vz_t)], \quad t=1,\ldots,T$
    \UNTIL convergence
    \STATE {\bfseries Return:} \, $\vJ_{1:T}, \vh_{1:T}$
 \end{algorithmic}
 \end{algorithm}

\subsection{Hyperparameters using log-marginal likelihood}
Discussed in the main text, the ELBO (represented below), for the models considered can be recast as
\begin{align}
	\mathcal{L}(\priorHypers) =& \sum_{t} \E_{q(\latents_t)} \log \frac{p(\vy_t \mid \latents_t)}{p(\auxLikelihood_t \mid \latents_t)} + \sum_{t} \log p(\auxLikelihood_t \mid \auxLikelihood_{1:t-1})
\end{align}
We see that, with parameters of the variational approximation fixed, only the log-marginal-likelihood of pseudo observations contributes to the gradient with respect to hyperparameters.  Then,
\begin{align}
    \nabla_{\priorHypers} \mathcal{L} &= \nabla_{\priorHypers} \sum \log p(\auxLikelihood_t \mid \auxLikelihood_{1:t-1})\\
    &= \nabla_{\priorHypers} \sum \log\left( \int p(\auxLikelihood_t \mid \vz_t^S) p(\vz^S_t \mid \auxLikelihood_{1:t-1}) \right)\\
    &= \nabla_{\priorHypers} \sum \log \Big(\int \mathcal{N}(\auxLikelihood_t \mid \vz_t, \auxCovariance_t)\mathcal{N}(\vz_t^S \mid \vm_t^{-}, [\vP_t^{-}]^{-1}) d\vz_t \Big)\\
    &= \nabla_{\priorHypers}\Big( -\tfrac{1}{2} (\auxLikelihood_t - \vH\vm_t^{-})^\top \vR_t^{-1} (\auxLikelihood_t - \vH \vm_t^{-}) - \tfrac{1}{2} \log |\vR_t|\Big)
\end{align}
where $\vR_t = \vH \vP_t^{-} \vH^\top + \auxCovariance_t$, and $\vm_t^{-} = \vA_{\priorHypers} \vm_{t-1}$, with $\vP_t^{-} = \vA_{\priorHypers} \vP_{t-1} \vA_{\priorHypers}^\top + \vQ_{\priorHypers}$ are the predictive means and covariances at time $t$, computed using the filtering step statistics; they depend on the kernel hyperparameters through the transition matrix, $\vA$, and the state noise $\vQ$.
\begin{figure*}[htbp]
    \centering
    \includegraphics[width=\textwidth]{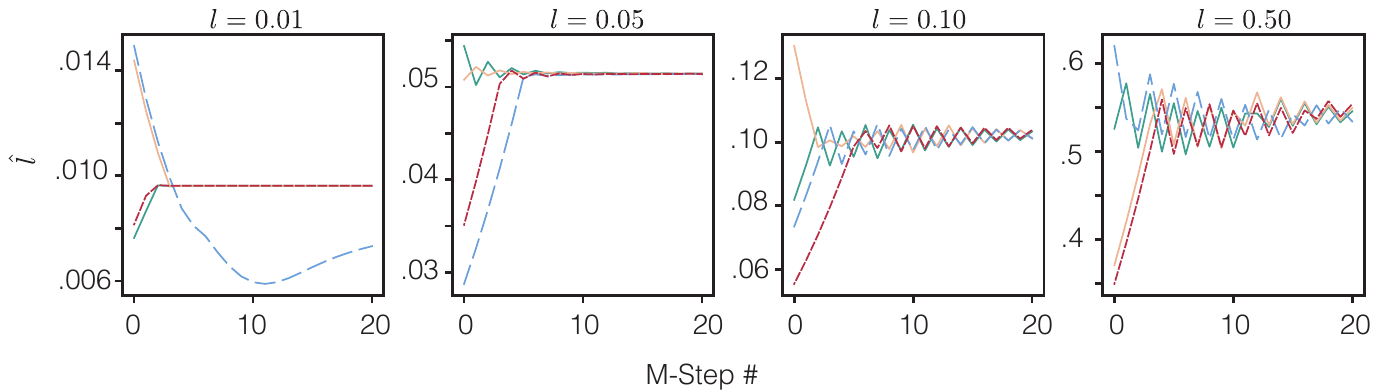}
    \caption{Optimal length scale estimated from the data during the M-steps.  We initialized the prior length scale randomly (4 realizations) about the true value to see how optimization performed. In most cases, the hyperparameters converged with a relatively small number of M-steps. The oscillation is likely attributed to the fact that we cap the gradient within 10-25\% of current value at each optimization step.}
    \label{fig:app:length_scale_optimization}
\end{figure*}
In order to verify hyperparameter optimization, we draw spike trains whose intensity are modulated by a \Matern 3/2 GP while varying the kernel lengthscale.  We use variational EM to estimate the hyperparameters of the prior as shown in Fig.~\ref{fig:app:length_scale_optimization}. 

\subsection{Poisson likelihood gradients}
For the update to the first natural parameter we have
\begin{align}
    \vh_k &= (1-\alpha) \vh_{k-1} + \alpha\vh_0\nonumber\\
    &\hspace{1em} + \alpha \nabla_{\vm(\naturalVariationalParams)} \sum_t \E_{q(\vz_t \mid \naturalVariationalParams)} \log p(\vy_t \mid \vz_t)
\end{align}
letting $l_{n,t} =  \E_{q(\vz_t \mid \naturalVariationalParams)} \log p(y_{n, t} \mid \vz_t)$, we have through the chain rule and natural parameter properties~\cite{wilkinson_spatiotemporal_vgp_2021} that
\begin{align}
    \nabla_{\vm(\naturalVariationalParams)} l_{n,t} &= \nabla_{\vm} l_{n,t} - 2 \nabla_{\vP} l_{n,t} \vm_t\\
    &= \vC_n \left(y_{n, t} - \Delta r_{n, t}\right) + \Delta r_{n, t} \vC_n \vC_n^\top \vm_t 
\end{align}
where $r_{n,t} = \exp(\vC_n^\top + \vb_n + \tfrac{1}{2} \vC_n^\top \vP_t \vC_n)$.  For the update to the second natural parameter we have
\begin{align}
    \vJ_k &= (1-\alpha) \vJ_{k-1} + \alpha \vJ_0\nonumber\\
    &\hspace{3em}+\alpha \nabla_{\meanVariationalParamsTwo(\naturalVariationalParams)} \sum_t \E_{q(\vz_t \mid \naturalVariationalParams)} \log p(\vy_t \mid \vz_t)
\end{align}
where
\begin{align}
    \nabla_{\meanVariationalParamsTwo(\naturalVariationalParams)} l_{n,t} &= \nabla_{\vP}\left(y_{n,t} \vC_n^\top \vm_t - \Delta r_{n,t}\right)\\
    &= -\tfrac{1}{2}\Delta \vC_n \vC_n^\top r_{n,t}
\end{align}

\section{Latent process velocity}\label{section:velocity}
A by-product of inference using the SSM representation of \HidaMatern GPs is that all mean square derivatives of the latent trajectories are inferred \textit{for free}.
Those mean square derivatives are equal to the sample path derivatives with probability one~\cite{Doob1990} and may reveal useful information about latent trajectories inferred.
One possibility, is using the mean square derivatives to probabilistically interpret the speed and acceleration at which neural trajectories evolve.
In the context of neural dynamics this can be useful; as an example, the velocity of latent neural dynamics is an interesting metric that may be used to substantiate or disprove certain hypothesis of neural computation~\cite{Sohn2019}.
While similar conclusions could possibly be drawn through strategies like finite differences, they would not provide a calibrated measure of uncertainty.  We examine this in the experiments section.

\section{PLDS with extended state-space}
\begin{figure*}[htbp]
    \centering
    \includegraphics[width=\textwidth]{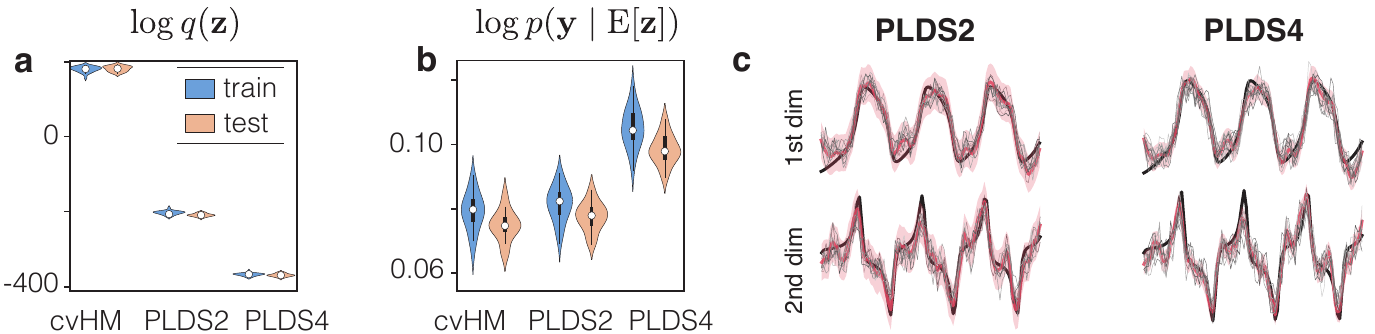}
    \caption{\CVIHMGP vs PLDS: For observations that read out $L$ latent variables, the dimension of \CVIHMGP's latent space will be $M=\sum M_l$ when the $l^{\text{th}}$ latent variable is modeled by an $M_l$-ple \HidaMatern GP. We compare \CVIHMGP with $M_1=2$ and $M_2=2$ to PLDS with latent dimensionality of 2 and 4. (\textbf{a}) log-likelihood of the `ground truth' under the posterior for \CVIHMGP and PLDS2 and PLDS4 (2 and 4 respectively denote the dimensionality of the latent space imposed).  (\textbf{b}) bits-per-spike (\textbf{c}) comparison of posterior for PLDS2 and PLDS4; PLDS4 has been projected down from 4 dimensions to 2.}
    \label{fig:app:plds_expanded_latent}
\end{figure*}

Our comparisons with PLDS in the paper used the same latent dimensionality.  However, the state-space dimensionality of \CVIHMGP will necessarily be inflated due to the need to propagate mean square derivatives of the latent processes.  In ~Fig.~\ref{fig:app:plds_expanded_latent} we show the result of an additional experiment where PLDS is also given an extended state-space. 

\section{Non-conjugate Gaussian observation example}
\begin{figure*}[htbp]
    \centering
    \includegraphics[width=\textwidth]{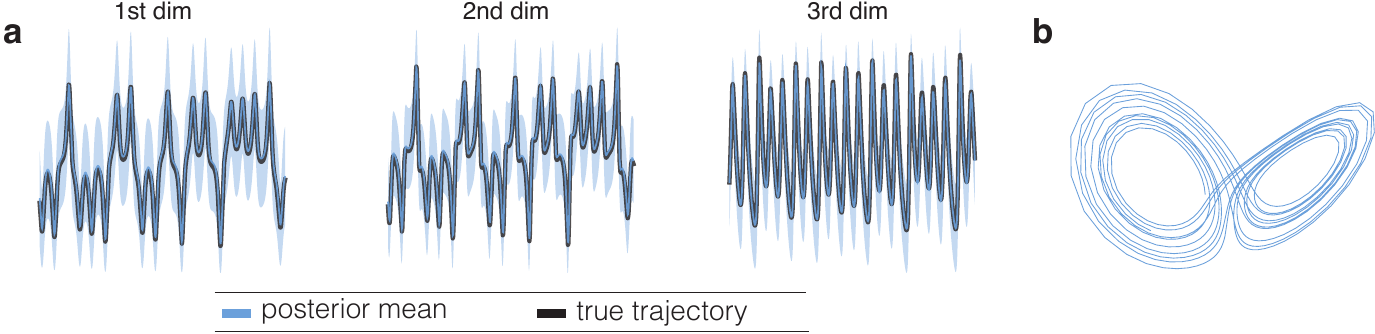}
    \caption{Nonlinear readout example: (\textbf{a}) \CVIHMGP infers accurate latents for the Lorenz system, observations are nonlinear readouts of a linear projection of the latent state; shading indicates the 95\% credible interval (\textbf{b}) mean inferred trajectories plotted in 3D}
    \label{fig:app:lorenz_nonlinear_readout}
\end{figure*}

To demonstrate the ability of \CVIHMGP to handle variety of non-conjugate cases, we consider nonlinear Gaussian observations readout from the latent state according to the following generative model
\begin{align}
    &\vz_{l,1:T} \sim \mathcal{GP}(0, k_l(\tau)) \qquad &l=1,\ldots,L\\
    &\vy_{n,t} \sim \mathcal{N}(\vy_{n, t} \mid g(\vz_t), \sigma_n^2) \qquad &n=1,\ldots,N\\
    &g(\vz_t) = \exp(\vC_n^\top \vz_t + \vb_n)
\end{align}
Following the prescription earlier, the only adjustments we need to make to infer the latent trajectories are to calculate the derivatives of the expected log-likelihood under our variational approximation.  Doing so, we have for $l_{n, t} = \E_{q(\vz_t \mid \naturalVariationalParams)} \log p(\vy_{n, t} \mid \vz_t)$, the gradients for the mean and covariance are:
\begin{align*}
    &\nabla_{\vm_t} l_{n,t} = -\frac{1}{\sigma^2_n}  \vC_n r_{n, t}\left[  r_{n,t} - \vy_{n, t}  \right]\\
    &\nabla_{\vP_t} l_{n,t} = -\frac{1}{\sigma^2_n}\vC_n\vC_n^\top r_{n,t} \left[  r_{n,t} - \vy_{n, t}  \right]
\end{align*}
Fig.~\ref{fig:app:lorenz_nonlinear_readout} shows the result of inference under this generative model when the true latent trajectories are generated according to the Lorenz system.  In this case, we were able to analytically calculate the expected log-likelihoods; in general this may not always be possible, in which case we can resort to a sampling scheme to approximate the intractable expectations required for CVI.

\section{Comparisons against other Markovian GPs}\label{app:markovian_gp_comparisons}
In the main text, comparisons were made against models aimed at finding structured representations of data where prior beliefs are specified according to GPs (e.g. PLDS, GPFA, and vLGP).  Here, we compare approximate GP regression with \HidaMatern kernels versus popular Markovian GPs~\cite{sarkka_hartikainen_ssms_2010,solin_infinite_horizon_gp,wilkinson_spatiotemporal_vgp_2021}; we choose two common baselines that use Poisson likelihoods from~\citet{wilkinson2020state}.

\begin{wrapfigure}[20]{r}{0.4\textwidth}
    \vspace{-5pt}
    \centering
    \includegraphics[width=0.4\textwidth]{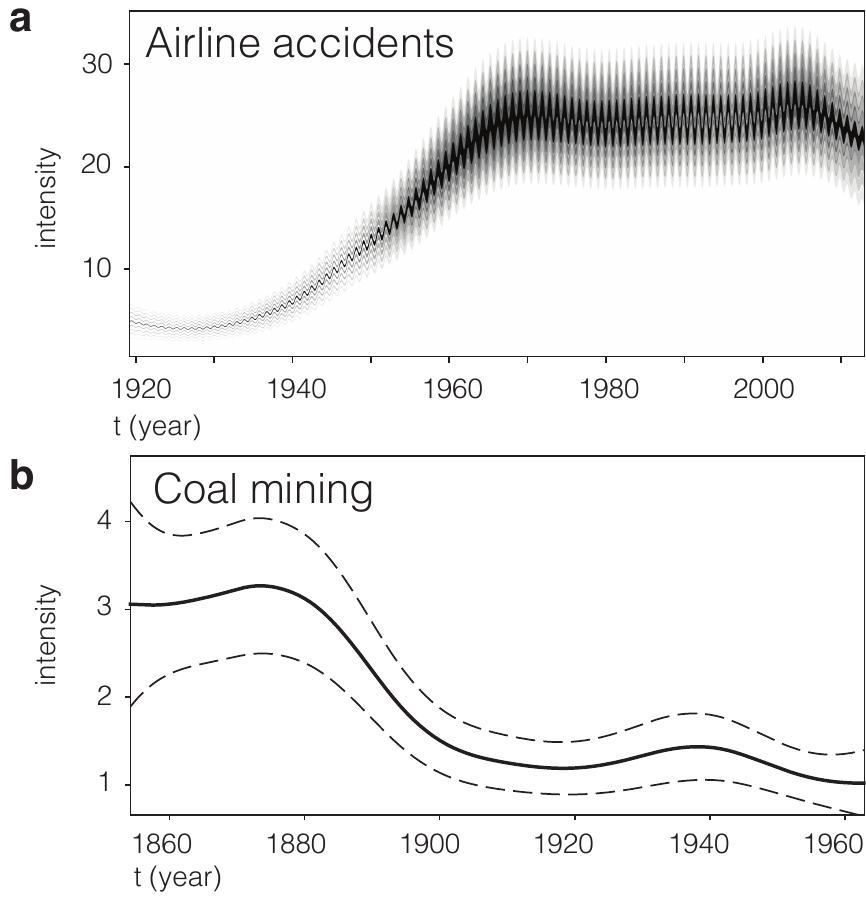}
    \caption{\textbf{a)} Mean airline accidents intensity and the 95\% credible interval.  \textbf{b)} Mean coal mining explosion accidents and 95\% credible interval.}
    \label{fig:app:aircraft_dataset}
\end{wrapfigure}

\begin{figure*}[htbp]
    \centering
    \includegraphics[width=0.67\textwidth]{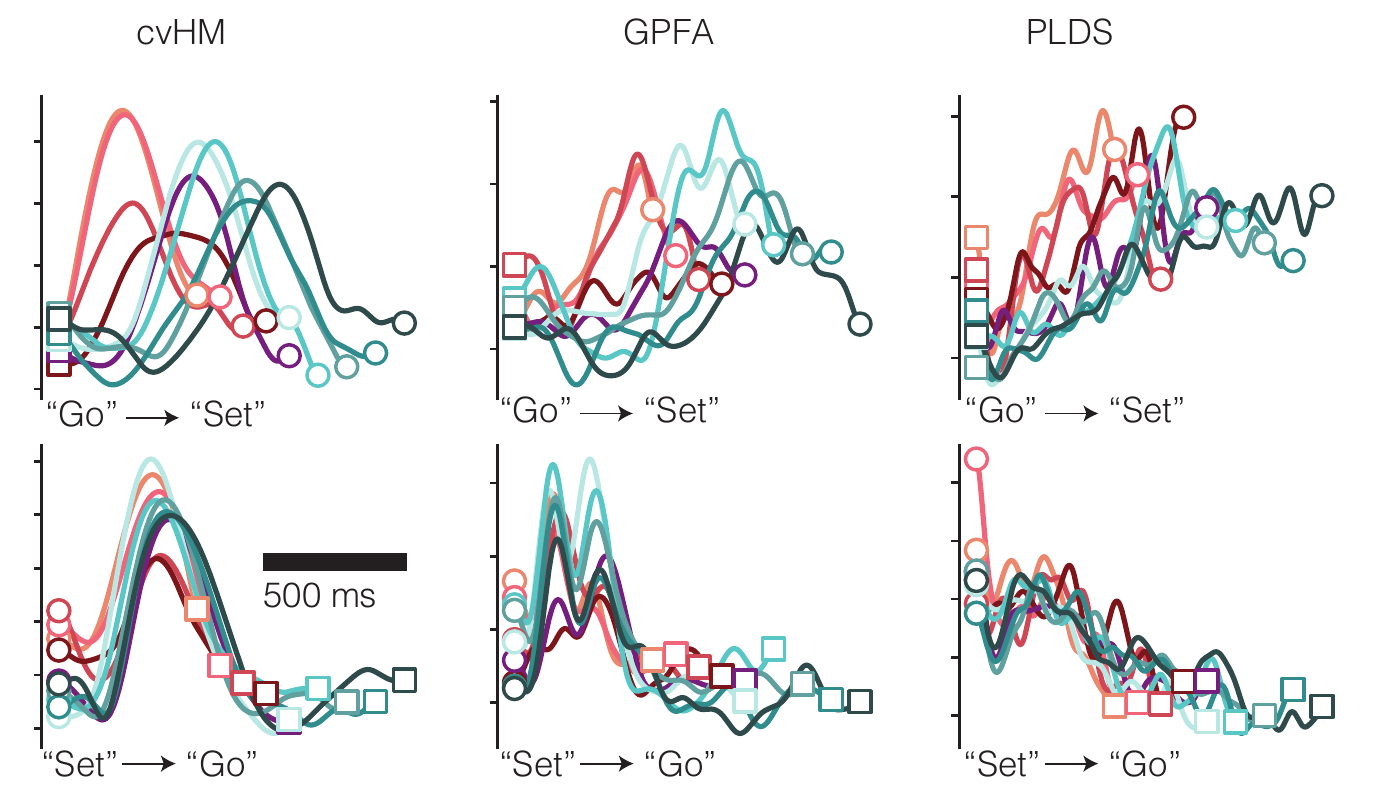}
    \caption{DMFC-RSG: \textbf{eye-left} condition. Similar to the eye-right condition presented in the main text, we can see that \CVIHMGP and GPFA recover latent trajectories which have peak speeds that decrease with respect to increasing intervals within the same prior.  This effect is harder to see in the trajectories inferred by PLDS.}
    \label{fig:app:dmfc_eye_left}
\end{figure*}

\begin{figure*}[htbp]
    \centering
    \includegraphics[width=0.67\textwidth]{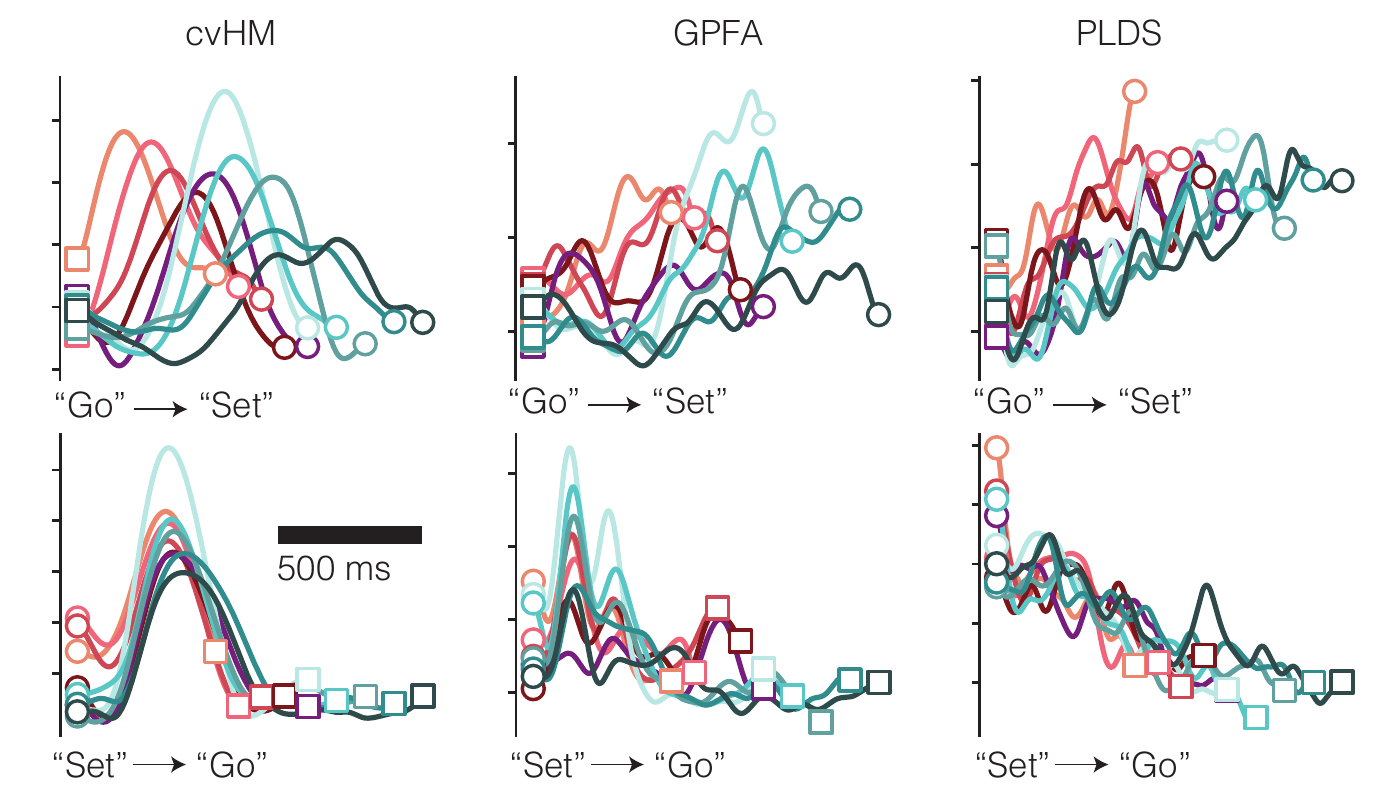}
    \caption{Condition: \textbf{eye-right}.  Neural trajectory velocities for the eye-right condition as presented in the main paper.  Again, PLDS seems to have inferred trajectories with an effective lengthscale that is too small.}
    \label{fig:app:dmfc_eye_right}
\end{figure*}

\begin{figure*}[htbp]
    \centering
    \includegraphics[width=0.67\textwidth]{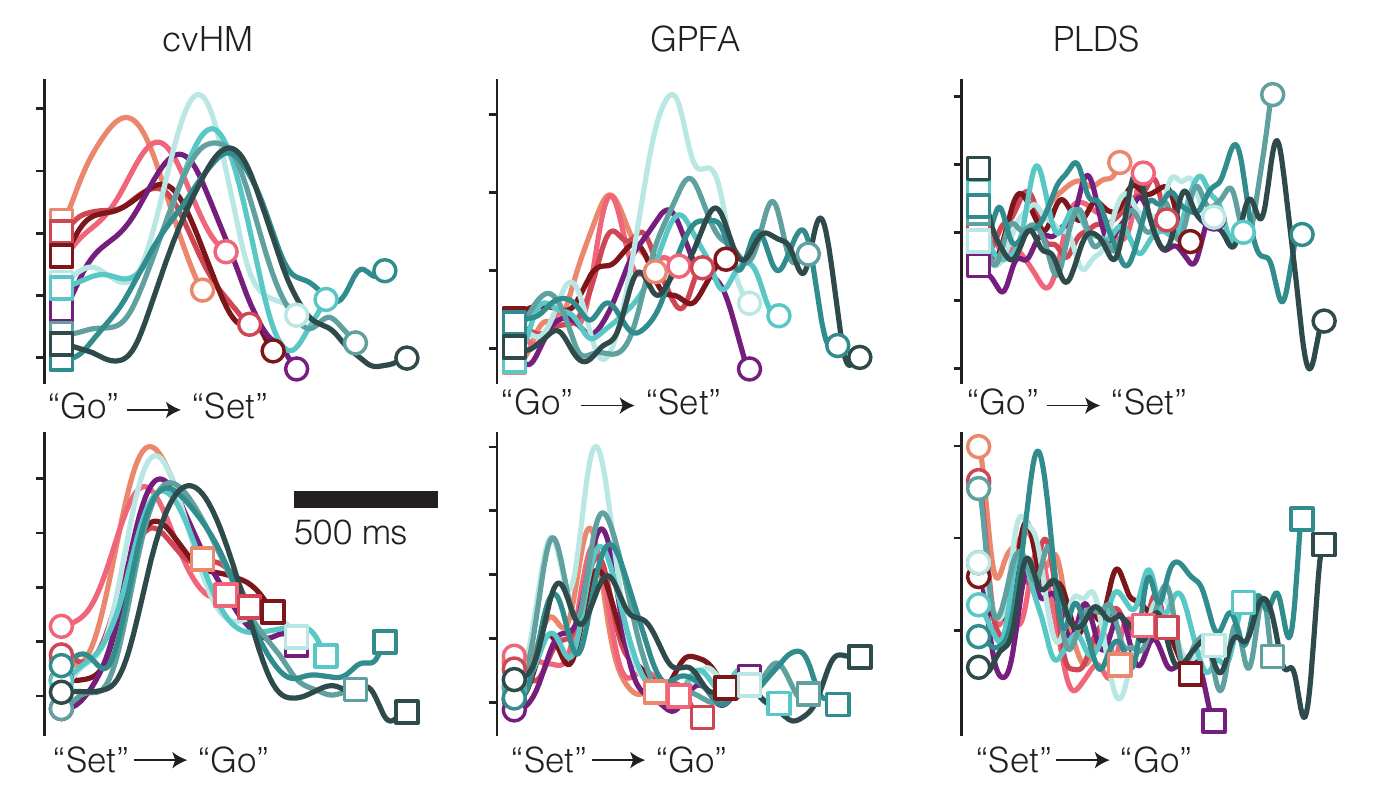}
    \caption{Condition: \textbf{hand-left}.  In comparison to the conditions requiring an eye saccade to indicate interval predictions, trajectories seem to end at higher speeds as seen in `Go' to 'Set'}
    \label{fig:app:dmfc_hand_left}
\end{figure*}

\begin{figure*}[htbp]
    \centering
    \includegraphics[width=0.67\textwidth]{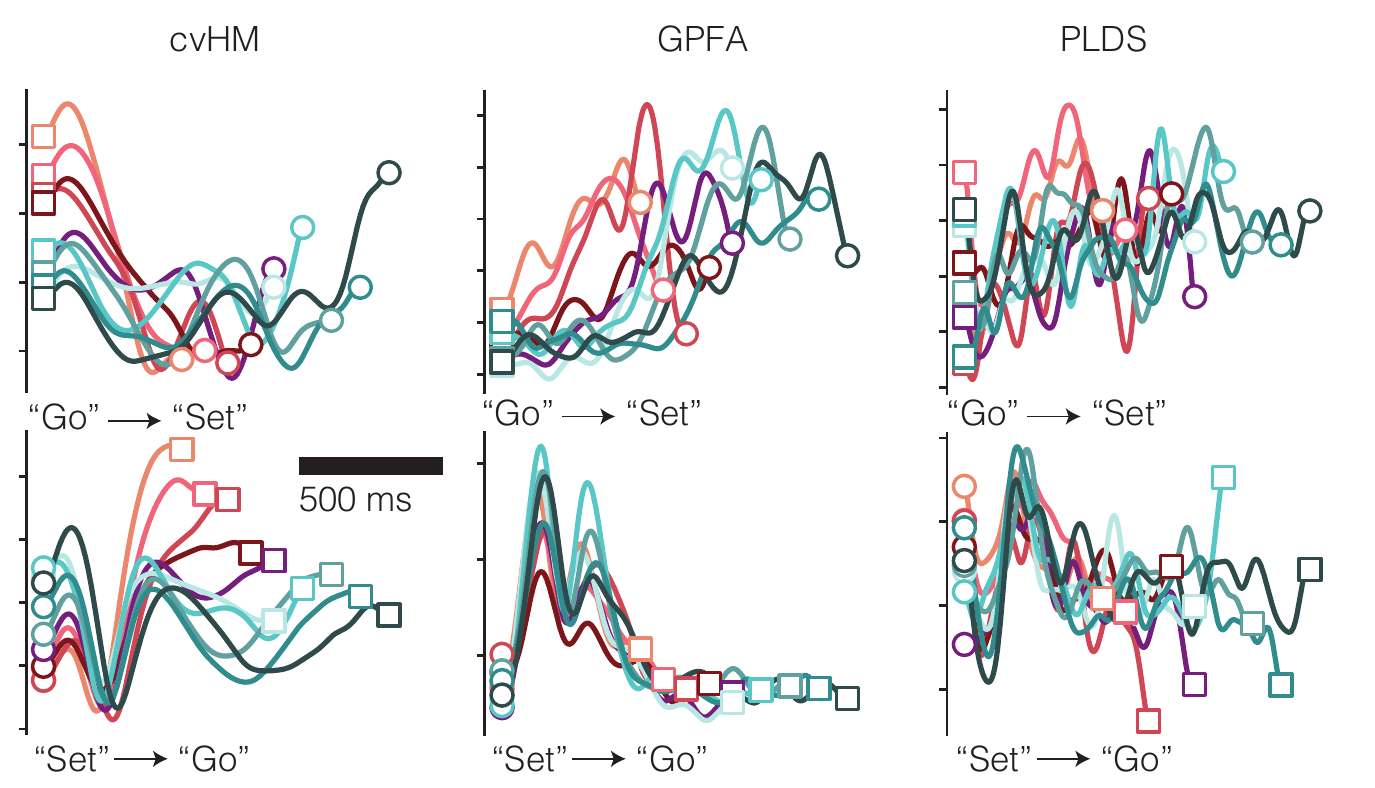}
    \caption{Condition: \textbf{hand-right}.  \CVIHMGP inferred trajectories for this condition are distinctly different than those it inferred in the three other conditions; the 'bump' like characteristic of the speed exists in short prior but not so much the long prior; additionally the average trajectory speed dips from its starting value for all interval times.}
    \label{fig:app:dmfc_hand_right}
\end{figure*}

\textbf{Coal dataset}\hspace{0.5em} The coal mining dataset reports the 191 coal mining explosions that killed 10 or more people in Great Britain from 1851 to 1962~\cite{wilkinson2020state}.  In order to make a fair comparison, we parameterize a GP similar to~\cite{wilkinson_spatiotemporal_vgp_2021} by using a second order \HidaMatern GP, and a Poisson likelihood with canonical link function so that the generative model is
\begin{align}
    p(\vz_{1:T}) &= \mathcal{N}(\vz_{1:T}\mid\boldzero, \vK_{TT})\\
    p(y_t\mid z_t) &= \text{Poisson}\left(y_t\mid\Delta\exp(z_t + b)\right)
\end{align}
where $\Delta$ is the bin size and $b$ is the bias.  We use the Whittle likelihood and bidirectional information filtering and calculate a 10 fold cross validated negative predictive log marginal likelihood of $0.955 \pm 0.16$, whereas the Markovian GP baselines used in~\cite{wilkinson2020state} achieve a 10 fold cross validated log marginal likelihood of $0.922 \pm 0.11$, when using the same data splits. \CVIHMGP performs worse on this dataset, possibly because the bias introduced by the Whittle approximation is significant in this low data regime.  We show the inferred posterior intensity in Fig.~\ref{fig:app:aircraft_dataset}\textbf{b}. 

\textbf{Aircraft accidents dataset.}\hspace{0.5em} The aircraft accidents dataset is another dataset that is well modeled by a Poisson likelihood. For a fair comparison against the approximate GP regression methods reported in~\cite{wilkinson2020state},  the GP prior kernel is constructed as
\begin{align*}
    k(\tau) &= \sigma^2_1 k_{H,M}(\tau; \tfrac{5}{2}, 0, \rho_1) + \sigma^2_2 k_{H,M}(\tau; \tfrac{3}{2}, b_2, \rho_2) \\
    &\hspace{3.0em}+ \sigma^2_3 k_{H,M}(\tau; \tfrac{3}{2}, b_3, \rho_3)
\end{align*}
\CVIHMGP achieves a 10 fold cross validated negative predictive log marginal likelihood of $0.142 \pm 0.01$ on this dataset which is the same as results reported in~\cite{wilkinson2020state}.  As a consequence of this dataset being larger than the coal dataset, \CVIHMGP is able to achieve results on par with other Markovian GP baselines even though it uses an approximation of the Gaussian log-marginal likelihood. The inferred posterior intensity is plotted in Fig.~\ref{fig:app:aircraft_dataset}\textbf{a}. 
\section{Ablation studies}\label{app:ablation_studies}
We tease apart some factors of \CVIHMGP that may make it preferable over a traditional/standard implementation that does not incorporate use of the Whittle likelihood or bidirectional information filtering with a few short experiments.

\textbf{Whittle likelihood and latent variable recovery}\hspace{0.5em}In the main paper, it was empirically shown that the Whittle approximation may produce a favorable loss landscape for hyperparameter optimization.  Given that variational EM is an iterative algorithm, we would expect that improvements in hyperparameter estimates lead to improvements in the variational approximation to the posterior of latent trajectories.  

Motivated by this, we consider a simple experiment where Poisson spiking is generated by a latent state simulated from a noisy Van der Pol oscillator like in Fig.~\ref{fig:vanderpol}.  Over 3 random seeds, we generate datasets of different number of trials, then measure the average log probability of the inferred latent trajectories when using the Whittle approximation or exact log-marginal likelihood -- the results are reported in Table~\ref{table:whittleLatentStateRecovery}.
\begin{table}[t]
	\sisetup{
		table-align-uncertainty=true,
		separate-uncertainty=true,
	  }
	  \renewrobustcmd{\bfseries}{\fontseries{b}\selectfont}
	  \renewrobustcmd{\boldmath}{}

	\begin{center}
	\begin{small}
	\begin{sc}
	\begin{tabular}{@{}l|rrr@{}}
			& \multicolumn{3}{c}{No. pts. $(\times 1000)$} \\
		\toprule
		Likelihood & 1 & 10 & 25 \\ \midrule
        exact    & -44.5 $\pm$ 14.4         & -1.55 $\pm$ 0.72          & -0.557 $\pm$ 0.680         \\
        whittle  & \bfseries{-8.5} $\pm$ \bfseries{3.52}      & \bfseries{-1.02} $\pm$ \bfseries{0.66}      & \bfseries{-0.408} $\pm$ \bfseries{0.121}   \\
		\bottomrule 
	\end{tabular}
	\end{sc}
	\end{small}
	\end{center}
	\caption{\textbf{Quantifying latent state recovery using Whittle versus exact log-likelihood}. In Fig.~\ref{fig::whittle_vs_lml} we highlighted the difference between the loss landscape generated by the exact log-likelihood versus the Whittle approximation; here, we explore the Whittle approximation's effect on latent state recovery in the approximate inference setting.  Numbers show the log-likelihood of the latent trajectories under the approximate posterior.}
	\label{table:whittleLatentStateRecovery}
\end{table}

\textbf{Information filtering improves lower floating point results}\hspace{0.5em}In this small experiment, we investigate how recovery of latent trajectories differs when using the information or covariance form of the Kalman filter.  Data is generated from Poisson observations with Van der Pol dynamics, similar to the previous experiment; over 5 random seeds, we draw 15 trials of length 1000, and then perform inference using either the information form of the Kalman filter, or the covariance form.  

For each seed, the average log probability of latent trajectories is calculated and plotted in Table~\ref{table:informationFilterFloatingPt}.  For 64 bit floating point, there is no advantage to using the information filter.  However, results for the covariance filter are consistently worse across each seed when the floating point precision is brought down to 32 bits.

\begin{table}[t]
	\sisetup{
		table-align-uncertainty=true,
		separate-uncertainty=true,
	  }
	  \renewrobustcmd{\bfseries}{\fontseries{b}\selectfont}
	  \renewrobustcmd{\boldmath}{}

	\begin{center}
	\begin{small}
	\begin{sc}
    \begin{tabular}{cllllll}
        \multicolumn{1}{l}{}             &                                  & \multicolumn{5}{c}{Seed No.}                                                                                          \\ \cline{3-7} 
        \multicolumn{1}{l}{Floating Pt.} & Filter                           & \multicolumn{1}{c}{1} & \multicolumn{1}{c}{2} & \multicolumn{1}{c}{3} & \multicolumn{1}{c}{4} & \multicolumn{1}{c}{5} \\ \hline
        \multirow{2}{*}{32 bit}          & \multicolumn{1}{l|}{covariance}  & -52,035,944           & -50,464,916           & -52,000,688           & -52,378,684           & -51,895,720           \\
                                         & \multicolumn{1}{l|}{information} & -52,030,328           & -50,463,748           & -51,997,296           & -52,370,120           & -51,893,184           \\ \hline
        \multirow{2}{*}{64 bit}          & \multicolumn{1}{c|}{covariance}  & -51,920,061           & -50,352,983           & -51,884,765           & -52,261,973           & -51,780,153           \\
                                         & \multicolumn{1}{c|}{information} & -51,920,061           & -50,352,983           & -51,884,765           & -52,261,973           & -51,780,153          
        \end{tabular}
	\end{sc}
	\end{small}
	\end{center}
	\caption{\textbf{Effect of filter type on latent variable recovery}. The average log-probability of latent trajectories for each seed is plotted as a function of the type of filter used, and the floating point precision.  Only when dropping the floating point precision down to 32 bits do we see the benefit of using the information filter over the standard covariance filter.}
	\label{table:informationFilterFloatingPt}
\end{table}

\newpage

\section{DMFC-RSG figures}\label{section:dmfc_figures}

\end{document}